\documentclass[a4paper]{article}
\usepackage[utf8]{inputenc}
\usepackage[textwidth=17.0cm,textheight=25cm]{geometry}
\usepackage{amsmath,amsfonts,amssymb,color,graphicx,epsf,subfigure}
\usepackage{authblk}
\usepackage[toc,page]{appendix}

\newcommand{\eq}[1]{(\ref{eq:#1})}  
\newcommand{\x}{{\boldsymbol{x}}}  
\newcommand{\kk}{{\boldsymbol{k}}}  
\newcommand{\Op}{\mathcal{O}}  
\newcommand{\ii}{\textrm{i}\,}  
\newcommand{\Tr}{\text{Tr}}  
\newcommand{\oin}{\omega_\text{in}}  
\newcommand{\oout}{\omega_\text{out}}
\newcommand{\op}{\omega_+}
\newcommand{\om}{\omega_-}
\newcommand{\mins}{{m_\text{in}}_{}}
\newcommand{\mout}{{m_\text{out}}_{}}
\newcommand{\ins}{\text{in}}
\newcommand{\out}{\text{out}}
\newcommand{\ket}[1]{|#1\rangle}
\newcommand{\bra}[1]{\langle#1|}

\newcommand{\dt}{\delta t}

\title{\bf Evolution of Complexity following a quantum quench\break in free field theory}
\author[1,2]{Daniel W. F. Alves\footnote{Email: \texttt{dwfalves@gmail.com}}}
\author[3]{Giancarlo Camilo\footnote{Email: \texttt{gcamilo@iip.ufrn.br}}}
\affil[1]{\footnotesize São Paulo State University (UNESP), Institute for Theoretical Physics (IFT) \break R. Dr. Bento T. Ferraz 271, Bl. II, 01140-070, Sao Paulo, SP, Brazil}
\affil[2]{\footnotesize Center for Quantum Mathematics and Physics, Department of Physics, University of California \break Davis CA 95616 USA}
\affil[3]{\footnotesize International Institute of Physics, Federal University of Rio Grande do Norte\break Campus Universit\'ario, Lagoa Nova, 59078-970, Natal, RN, Brazil}
\date{\vspace{-5ex}}

\begin{document}

\maketitle

\begin{abstract}
Using a recent proposal of circuit complexity in quantum field theories introduced by Jefferson and Myers, we compute the time evolution of the complexity following a smooth mass quench characterized by a time scale $\delta t$ in a free scalar field theory. We show that the dynamics has two distinct phases, namely an early regime of approximately linear evolution followed by a saturation phase characterized by oscillations around a mean value. The behavior is similar to previous conjectures for the complexity growth in chaotic and holographic systems, although here we have found that the complexity may grow or decrease depending on whether the quench increases or decreases the mass, and also that the time scale for saturation of the complexity is of order $\delta t$ (not parametrically larger).
\end{abstract}

\section{Introduction}\label{intro}

Entanglement, the most distinct trait of quantum mechanics, has proven to be a key concept in many different areas of physics. In particular, it has recently become a standard term in the high-energy physics vocabulary due mainly to the AdS/CFT correspondence, the Ryu-Takayanagi formula \cite{Ryu:2006bv} and its generalization \cite{Hubeny:2007xt}. Many previous works have explored these connections (see, e.g., \cite{VanRaamsdonk:2016exw} for a review) with the aim of elucidating the basic mechanism of the AdS/CFT correspondence and space-time emergence.

Some questions remain unanswered though, in particular how the behavior of the interior of black holes (which is a region often inaccessible to Ryu-Takayanagi surfaces) in asymptotically AdS space-times is encoded in the dual CFT. In particular, by a classical gravity computation, it is possible to show that the volume of the Einstein-Rosen bridge connecting the two sides of an eternal AdS black hole grows linearly in time, which raises the question of what quantity this corresponds to in the dual CFT \cite{Susskind:2014moa}. 

As we will explain below, it has been proposed that the computational complexity in the field theory, or complexity for short,  could be the quantity whose growth is the dual to that of the Einstein-Rosen bridge in the gravity side. Complexity is a concept originally introduced in the field of classical computation \cite{Arora:2009:CCM:1540612}, though it is also extensively studied in quantum information.
The basic idea is to consider some reference state $\ket{\Omega}$ and some target state $\ket{\psi}$, as well as a set of quantum operators or gates, $\Op_{I}$. The complexity of $\ket{\psi}$ is then the length of the minimal circuit connecting it to $\ket{\Omega}$ by successive applications of the operators $\Op_{I}$. To properly define \lq\lq length\rq\rq and \lq\lq minimal\rq\rq, we need a notion of distance, or cost, associated to a given circuit, that in principle we are free to define. The definitions of cost and the choice of $\Op_{I}$ should be made as to capture the intuition that the gates are $\it{easy}$ to apply. For instance, the operators should act on only a few degrees of freedom at a time. It is also possible to study the complexity of quantum operators themselves \cite{Yang:2018nda} instead of quantum states, but in this paper we focus on the latter.

Two conjectures which aim to connect the complexity of the boundary field theory to geometric constructs in the bulk space-time have been proposed, the so called \lq\lq complexity $=$ action\rq\rq \cite{Brown:2015bva} and \lq\lq complexity $=$ volume\rq\rq \cite{Stanford:2014jda} proposals.  In these works, it was argued that these objects in the bulk indeed grow as expected. The bulk computation of the evolution of holographic complexity was put on firmer grounds by \cite{Carmi:2017jqz}, and it has also been subject of many recent works \cite{Swingle:2017zcd,Alishahiha:2018tep,Alishahiha:2017hwg,Cano:2018aqi,Chapman:2016hwi,Couch:2017yil,Fu:2018kcp,Ghodrati:2017roz,An:2018xhv,Kim:2017qrq}. For instance, in \cite{Moosa:2017yvt}, the holographic complexity following a quantum quench was calculated by modeling the quench with an AdS-Vaidya spacetime and it was found that the complexity evolves approximately linearly.
 
To truly check the proposed correspondence, it is necessary to have a precise definition of complexity in quantum field theory in the first place. That was only put forward very recently \cite{Jefferson:2017sdb,Chapman:2017rqy,Khan:2018rzm,Hashimoto:2017fga,Yang:2017nfn}, where the authors defined and calculated the ground state complexity of free quantum field theories. Their approach was inspired by previous work by Nielsen \cite{2006Sci...311.1133N} where the task of finding the optimal circuit connecting two quantum states was turned into a geometric problem of finding geodesics in a curved space. Some similarities with the general relativistic computation of holographic complexity were reported in \cite{Jefferson:2017sdb}, and \cite{Chapman:2017rqy}, even though the theories with classical gravity duals are known to be very different, i.e., strongly coupled and with a large number of degrees of freedom.

The study of complexity in quantum field theory might also prove useful for subjects unrelated to holography. For instance, in \cite{Roberts:2016hpo}, a lower bound on the complexity of an ensemble of operators in lattice systems was obtained in terms of out-of-time order correlators. So, in this sense quantum chaos and complexity seem to be quantitatively related and it would be interesting to check if something similar happens in continuum QFT's. It has also been conjectured \cite{Brown:2017jil} that a universal second law of complexity should hold for operators in quantum chaotic systems (analogous to the second law of thermodynamics).  

In the present work, we use the definition of \cite{Jefferson:2017sdb} and, following a construction similar to the one done in \cite{Nozaki:2012zj} and \cite{Mollabashi:2013lya} in the context of cMERA, we calculate the complexity of circuits preparing the instantaneous state after a quantum quench and study its time evolution. Our basic aim here is to explore the definition \cite{Jefferson:2017sdb} in a dynamical, non-equilibrium, setting, to compare our results with holographic calculations, and to use them as a benchmark for future studies on interacting theories. The calculation becomes particularly simple if the so called $F_{2}$ complexity is used, since in this case the problem of finding the optimal circuit geometrizes to that of computing geodesics in a Riemannian curved space. 
We shall deal with mass quenches in free bosonic theory. By choosing to work with free field quenches, we keep the problem computationally tractable while still allowing for non-trivial dynamics. In this case, one should not expect thermalization in the usual sense of Gibbs ensemble, since the free scalar field is an integrable system. Nevertheless, it was shown in \cite{Sotiriadis:2014uza} that many such integrable systems do thermalize in a broader sense, approaching the so called Generalized Gibbs Ensemble (GGE). In fact, it has been shown in \cite{Alves:2017fjk} that after the quench the entanglement entropy in momentum-space agrees with the prediction from a GGE, with the density matrix after the quench being explicitly computed in terms of the quench data.

The paper is organized as follows. We begin with a brief review of smooth mass quenches in free scalar theories in Section {\bf\ref{mass_quenches_in_free_scalar_fields}}. We then discuss the definition of circuit complexity in QFT as proposed in \cite{Jefferson:2017sdb} in Section {\bf\ref{cpxt_in_qft}}. In Section {\bf\ref{introducing_the_circuit}}, we introduce our circuit by defining the reference and target states, as well as the elementary gates. The complexity is then calculated in Section {\bf\ref{time_evolution_if_the_cpxt}}, the final result being given by \eq{cpxt_finalanswer}. The final remarks appear in Section {\bf\ref{conclusion}}. In Appendix {\bf A}, we prove the equivalence of Schr\"odinger and Heisenberg picture calculations of the complexity, which is an important step of the calculation.

\vspace{.5cm}
\noindent{\bf Note added:} After the paper appeared online we have been informed of an upcoming paper \cite{PAWEL} containing a very similar analysis.

\section{Mass quenches in free scalar fields}\label{mass_quenches_in_free_scalar_fields}

We begin with a quick review of the approach introduced in \cite{Das:2014hqa} (the reader is referred to \cite{Alves:2017fjk} for details) to study mass quenches in a free bosonic quantum field theory. The action describing the system is 
\begin{equation}\label{eq:scalarfieldaction}
I=\frac{1}{2}\int d^{d}x\,\big(\eta^{\mu\nu}\partial_{\mu}\phi\partial_{\nu}\phi-m(t)^{2}\phi^{2})\,,
\end{equation}
where the mass profile $m(t)$ asymptotes to constants $\mins\equiv m(-\infty)$
and $\mout\equiv m(+\infty)$ at early and late times, respectively, so that asymptotic states exist before and after the quench. We shall focus on $d=2$ for simplicity. This problem is equivalent to a standard quantum scalar field of constant mass $m_0$ placed in a cosmological background and therefore can be understood using intuition from quantum field theory in curved spacetimes \cite{Alves:2017fjk}. We focus on the particular choice 
\begin{equation}\label{eq:massquenchtanh}
 m(t)^2 = \frac{1}{2}\left(\mout^2+\mins^2\right)+\frac{1}{2}\left(\mout^2-\mins^2\right)\tanh \frac{t}{\delta t}\,,
\end{equation}
a smoothened version of a step function connecting the initial and final masses where the nontrivial time dependence happens roughly within the scale $\delta t$, since in this case exact solutions are known for the classical mode functions. 

There are two types of mode functions, the so-called in- and out-modes $\phi^\text{in/out}_{k}(t,x)=(2\pi)^{-1/2}e^{\ii kx}\chi^\text{in/out}_{k}(t)$, depending on how we choose to fix the boundary conditions. They are given by 
\begin{subequations}\label{eq:out_modes}
\begin{align}
\chi_{k}^\ins(t) &= \frac{1}{\sqrt{2\oin}}\,e^{-\ii\op t-\ii\om\dt\log(2\cosh\frac{t}{\dt})}{}_{2}F_{1}\left[1+\ii\om\dt;\ii\om\dt;1-\ii\oin\dt;\frac{1+\tanh\frac{t}{\dt}}{2}\right]\\
\chi_{k}^\out(t) &= \frac{1}{\sqrt{2\oout}}\,e^{-\ii\op t-\ii\om\dt\log(2\cosh\frac{t}{\dt})}{}_{2}F_{1}\left[1+\ii\om\dt;\ii\om\dt;1+\ii\oout\dt;\frac{1-\tanh\frac{t}{\dt}}{2}\right]\,,
\end{align}
\end{subequations}
where $_{2}F_{1}[...]$ is the hypergeometric function and 
\begin{equation}
\oin=\sqrt{k^{2}+{\mins}^{2}}\qquad\qquad \oout=\sqrt{k^{2}+{\mout}^{2}}\qquad\qquad\omega_{\pm}=\frac{1}{2}(\oout\pm\oin)\,.
\end{equation}
The in-mode satisfies $\lim_{t\rightarrow-\infty}\chi^\ins_k(t)\sim e^{-\ii\oin t}$ and therefore is adapted to observers at early times (at $t=-\infty$, before the quench starts), while the out-mode satisfies $\lim_{t\rightarrow\infty}\chi^\out_k(t)\sim e^{-\ii\oout t}$ and is adapted to post-quench observers at $t=+\infty$. 

Canonical quantization of the field $\phi$ then proceeds in the usual way. Both in and out-modes are equally good and the field can be expanded using either of them, i.e.,
\begin{equation}\label{eq:fieldinout}
\phi(t,x)=\int\frac{dk}{2\pi}\,e^{\ii kx}\big[a_{k}^{\ins/\out}\chi_{k}^{\ins/\out}(t)+a_{-k}^{\dagger\,\ins/\out}\chi_{-k}^{*\,\ins/\out}(t)\big]\,,
\end{equation}
To each set of modes (in/out) one can associate a corresponding vacuum state $\ket{0_{\ins/\out}}$, namely,
\begin{equation}\label{eq:vacuuminoutdef}
a_{k}^{\ins/\out}\,\ket{0_{\ins/\out}}=0\,,
\end{equation}
as well as particle excitations by acting with $a_k^{\dagger\,\ins/\out}$. They are of course different, being adapted either to early or late-time measurements only, though both correct. In fact, the two definitions are connected by a Bogoliubov transformation that is block-diagonal with respect to $k$ (i.e., it mixes only opposite momentum modes) of the type
\begin{equation}\label{eq:constraint_bogoliubov}
a_{k}^{\ins}=A_{k}\,a_{k}^{\out}+B_{k}\,a_{-k}^{\dagger \out}\,,
\end{equation}
where
\begin{subequations}\label{eq:bogoliubovcoeffs}
\begin{align}
A_{k} &= \left(\sqrt{\frac{\oout}{\oin}}\,\frac{\Gamma(1-\ii\oin\dt)\Gamma(-\ii\oout\dt)}{\Gamma(-\ii\op\dt)\Gamma(1-\ii\op\dt)}\right)^*\\
B_{k} &= \left(-\sqrt{\frac{\oout}{\oin}}\,\frac{\Gamma(1-\ii\oin\dt)\,\Gamma(\ii\oout\dt)}{\Gamma(\ii\om\dt)\,\Gamma(1+\ii\om\dt)}\right)^*\,
\end{align}
\end{subequations}
and $|A_k|^2-|B_k|^2=1$. From this it is straightforward to check that the in and out vacua are indeed physically different: $\ket{0_{\ins}}$, which contains no particles for an initial observer, is populated by $|B_k|^2$ particles of the out type (and vice-versa), which is the phenomenon of particle production by the quench.

\section{ Complexity in quantum field theory}\label{cpxt_in_qft}

We start by reviewing the definition of circuit complexity in quantum field theory as proposed in \cite{Jefferson:2017sdb}. As mentioned in the Introduction, to properly define complexity one first needs to specify a reference state $\ket{\Omega}$, a target state $\ket{\psi}$, and a set of elementary gates (quantum operators) $\{\Op_I\}$ to be used to build a quantum circuit connecting the them. 
The circuit then consists of a continuum set of operations parametrized by some parameter $s$ (we take $s\in[0,1]$ with no loss of generality) as
\begin{equation}
\ket{\psi}=\mathcal{P}\,e^{-\int_{0}^{1}Y^{I}(s)\Op_{I}\,ds}\,\ket{\Omega}\,,
\end{equation}
where the set of complex functions $Y^{I}(s)$ characterize a particular circuit roughly by deciding whether or not the gate $\Op_I$ is applied at~\lq\lq level $s$\rq\rq~of the circuit. As we shall see, in general there will be many distinct such functions preparing the same final state.
The path-ordering symbol $\mathcal{P}$ here ensures that the circuit is built in the direction of growing $s$, i.e., that operators are applied sequentially from $s=0$ to $s=1$. In the following it will be convenient to introduce 
\begin{equation}
U(s)=\mathcal{P}\,e^{-\int_{0}^{s}Y^{I}(s')\Op_{I}\,ds'}\,,\qquad\qquad U(0)=1\,
\end{equation}
such that the circuit becomes 
\begin{equation}
\ket{\psi}=U(1)\ket{\Omega}\,.
\end{equation}

Next, we need a notion of \emph{circuit depth} that will allow us to quantify the costs of applying different circuits to accomplish the same task of connecting a pair of reference and target states. The computational complexity of this task (or, equivalently, the complexity of the state $\ket\psi$ relative to $\ket\Omega$) will then be defined as the cost of the optimal circuit amongst all, i.e., the one with minimal depth. Given a circuit, or set of functions $\{Y^I(s)\}$, its depth can be quantified by any functional of the form 
\begin{equation}
{\mathcal D}\left[U\right]=\int_{0}^{1}F\left(Y^{I}(s),\partial_s{Y}^{I}(s)\right)ds\,
\end{equation}
with the cost function $F$ satisfying a list of reasonable physical requirements such as smoothness, positivity, and the triangle inequality (see \cite{2006Sci...311.1133N} for details). It remains unclear whether there is a preferred choice of $F$ and, for the sake of simplicity, we shall follow 
\cite{Jefferson:2017sdb} and focus here on the so called $F_{2}$ distance defined by $F=\sqrt{\delta_{IJ}\,Y^I\,(Y^J)^*}$, even though the strategy is exactly the same for any other distance measure. This means that, for our purposes, the circuit depth is defined by
\begin{equation}\label{eq:F_2_distance}
{\cal D}_{F_2}\left[U\right] = \int_{0}^{1}ds\,\sqrt{\sum_{I}\left|Y^{I}(s)\right|^2}
\end{equation}
and the corresponding complexity is simply
\begin{equation}
{\cal C}_{F_2} = {\cal D}_{F_2}\left[U_\text{optimal}\right] = \min_{\{Y^I\}} {\cal D}_{F_2}\left[U\right] \,.
\end{equation}

The main advantage of this approach, to be discussed in detail below, relies on the fact that the problem of minimizing ${\cal D}_{F_2}\left[U\right]$ to calculate the complexity can in principle be rephrased as a geometric problem of finding geodesics in a Riemannian manifold. Namely, for specific choices of reference and target states the set of gates $\Op_{I}$ needed to implement the task closes a Lie algebra and hence the circuits $U$ are elements of the corresponding Lie group. One can then cover the group manifold with coordinates $\x$ in such a way that quantum circuits become curves in this manifold; the curves are defined by $Y^{I}(s)=Y^I\left(\x(s),\partial_s\x(s)\right)$ and finding the optimal circuit amounts simply to finding the minimal length curve connecting the reference and target states. In particular, using the $F_{2}$ cost function above reduces the problem to that of calculating a geodesic in a Riemannian space\footnote{For more general cost functions $F$ we end up with Finsler geometries instead \cite{Jefferson:2017sdb}.} whose metric is given by
\begin{equation}\label{eq:line_element}
g_{\mu\nu}\frac{dx^{\mu}}{ds}\frac{dx^{\nu}}{ds}=\sum_{I}\left|Y^{I}(s)\right|^{2}\,.
\end{equation}

In the following we shall design quantum circuits preparing the time-dependent state following a quantum quench from a product state having no entanglement over spatial subregions. We will see that an infinite number of elementary gates is needed, each acting on a particular momentum mode $k$ of the quantum field.\footnote{If we work in a non-compact space, the $k$'s will form a continuous set, meaning that we would have an infinite and uncountable set of elementary gates. We shall work in a compact space, where the momentum label takes values on a discrete set and things are well-defined.}  These gates will be denoted by $\Op_{I,k}$ and the circuit by a set of functions $Y^{I}_{k}(s)$. Thus, we will write the circuit as 
\begin{equation}
U(s) = \mathcal{P}\,e^{\ii\int_0^s\,ds'\,\sum_k Y^I_{k}(s')\Op_{I,k}}\,.
\end{equation}
The sum runs over discrete values of $k$ for the field in the compact space and must be replaced by $V\int dk$ in the continuum limit, where $V$ is the spatial volume factor needed to make the exponent dimensionless. In our case, when it is time to take the continuum limit we shall ignore this factor and set $V=1$ for simplicity.

\section{Introducing the circuit}\label{introducing_the_circuit}
	\subsection{Reference and target states}

In this section, we specify the target and reference states that define our circuit. We shall choose our reference $\ket\Omega$ to be a product state, characterized by the absence of entanglement in any spatial subregion. A nice way to impose this condition (see \cite{Mollabashi:2013lya,Nozaki:2012zj,Jefferson:2017sdb} for similar work in the context of cMERA tensor networks) is by demanding the state to satisfy
\begin{equation}\label{eq:refstatedefS}
\text{Reference:}\qquad\left(\sqrt{M}\,\phi(x)+\frac{\ii}{\sqrt{M}}\pi(x)\right)\ket\Omega=0
\end{equation}
for some arbitrarily chosen mass scale $M$. It is straightforward to check that this definition implies \cite{Nozaki:2012zj}
\begin{subequations}
\begin{align}
\langle\phi(x)\phi(x')\rangle &= \delta(x-x')\\
\langle\pi(x)\pi(x')\rangle &= \delta(x-x')\,,
\end{align}
\end{subequations}
which indeed indicates that the state have no spatial entanglement. 

Our target state, on the other hand, will be time-dependent and shall be denoted $\ket{\psi(t)}$. We choose it to be the instantaneous state of the system at time $t$ following a mass quench that starts from the vacuum of the pre-quench Hamiltonian at $t=-\infty$, i.e., 
\begin{equation}\label{eq:targetstatedefS}
\text{Target:}\qquad\ket{\psi(t)} = E(t,-\infty)\,\ket{0_{\ins}}
\end{equation}
where $E$ is the time evolution operator. 

Notice that the states above are defined in the Schr\"odinger picture, which is why no time dependence shows up in the fields in \eq{refstatedefS} and the whole time dependence is contained in the target state $\ket{\psi(t)}$. Indeed, in the field representation $\ket{\{\phi(x)\}}$ (where $\pi(x)=-\ii\frac{\delta}{\delta\phi(x)}$) equation \eq{refstatedefS} can be seen as a functional differential equation defining the wave functional $\Omega[\phi(x)]\equiv\bra{\{\phi(x)\}}\Omega\rangle$, whose solution is a simple Gaussian function with width controlled by $M$. By discretizing the field into a chain of coupled harmonic oscillators this becomes precisely the Gaussian product state used by the authors of \cite{Jefferson:2017sdb} as the reference state with respect to which the free scalar field complexity is evaluated. Similarly, in the absence of a quench, i.e. the limit $\mins=\mout$, the time evolution in \eq{targetstatedefS} is trivial and the target state reduces simply to the vacuum of a free massive scalar field, which also corresponds to the choice of target state used in \cite{Jefferson:2017sdb}.\footnote{Actually even in the presence of the quench ($\mins\ne\mout$) our target state is still a Gaussian state of the same type used in \cite{Jefferson:2017sdb}, since the pre-quench vacuum is Gaussian and the time evolution operator (despite of its nontrivial time-dependence) is basically quadratic in the fields. However, in this case a different set of fundamental gates than the ones used in \cite{Jefferson:2017sdb} is needed to construct the time-dependent circuit in their setup (one including also the $\phi^2$ and $\pi^2$ generators, since they are the ones responsible for the nontrivial time evolution in the quench Hamiltonian and our target state is defined by time evolution). We thank Rob Myers for pointing this out. } 

Nevertheless, since the analysis of the exactly solvable quantum quench model of previous Section is done in the Heisenberg picture, it will prove more convenient to redefine our whole circuit and work in the Heisenberg picture instead. The physical intuition becomes slightly less obvious in this case though, since the roles are interchanged: the time dependence due to the quench now appears in the constraint defining the reference state (since the fields themselves are now time dependent), whereas the target state becomes constant. Anyway, the Heisenberg picture reference and target states are defined by
\begin{subequations}\label{eq:ref_state_t}
\begin{align}
\text{Reference:}&\qquad\left(\sqrt{M}\,\phi\left(x,t\right)+\frac{\ii}{\sqrt{M}}\,\pi\left(x,t\right)\right)\ket{\Omega\left(t\right)} = 0\\
\text{Target:}&\qquad\ket{\psi} = \ket{\psi(-\infty)} = \ket{0_{\ins}}\,.
\end{align}
\end{subequations}
or equivalently, using \eq{fieldinout}, \eq{vacuuminoutdef} and \eq{constraint_bogoliubov}, by the constraints
\begin{subequations}\label{eq:reftargetdef}
\begin{align}
\left[\alpha_{k}(t)\,a^\out_{k}+\beta_{k}(t)\,a_{-k}^{\dagger\out}\right]\ket{\Omega(t)}&=0\label{eq:reftargetdef1}\\
(A_{k}\,a_{k}^{\out}+B_{k}\,a_{-k}^{\dagger\out})\,\ket{\psi}&=0\,,\label{eq:reftargetdef2}
\end{align}
\end{subequations}
where $A$ and $B$ are the Bogoliubov coefficients \eq{bogoliubovcoeffs} and we have introduced
\begin{subequations}\label{eq:alphabetadef}
\begin{align}
\alpha_k(t)&\equiv \frac{1}{2}\left(\sqrt{M}\,\chi_{k}^\out(t)+\frac{\ii}{\sqrt{M}}\,\dot{\chi}_{k}^\out(t)\right)\\
\beta_k(t)&\equiv\frac{1}{2}\left(\sqrt{M}\,\chi_{-k}^{\out\,*}(t)+\frac{\ii}{\sqrt{M}}\,\dot{\chi}_{-k}^{\out\,*}(t)\right)\,.
\end{align}
\end{subequations}
Presenting the circuit in terms of the out modes is interesting since we are interested in studying time evolution of the complexity and, in particular, its late-time behavior after the quench has finished (a regime where the out modes are the most natural ones to use). For simplicity of notation, hereinafter we drop the superscript $^\out$ and convention that \emph{mode functions $\chi_k$ and creation/annihilation operators $a_k,a^\dagger_k$ carrying no superscript always refer to out.}

In principle it is not quite obvious that the circuit complexity will be the same in both Schr\"odinger and Heisenberg pictures, so in Appendix \ref{app:SH} we show explicitly that this is the case. Also, as discussed above, in the limit where $\mins=\mout$ (no quench) our setup is equivalent to the one studied in \cite{Jefferson:2017sdb}, so as consistency check we shall be able to recover their results using Heisenberg picture calculations.

Having defined the target and reference states, the next task is choosing a set of gates connecting them. From now on it is always implicit that we are working in the Heisenberg picture.

	\subsection{Choosing the gates}

The choice of gates used to build a circuit connecting the reference and target states is ambiguous, since there are not many requirements to constrain the options. In addition to being able to implement the desired task in a unitary way, we only require that the gates should act on the smallest possible number of degrees of freedom at each step, so that they constitute the most basic possible set of operations needed to construct the final state. This captures the intuition that the gates are \textit{easy} to apply, and that the complexity then measures how difficult it is to complete the task in the sense of counting the minimum number of elementary operations needed. In a lattice system, for instance, this can be achieved by using gates that only act on neighboring lattice sites, i.e., gates comprising a notion of real space locality. But this is not the only option, and indeed the gates we shall introduce are different: they are local in momentum space, in the sense that they only act on specific fixed-momentum subspaces of the full Fock space (recall that the Fock space of free theories factorizes into subspaces of fixed-momentum modes \cite{Alves:2017fjk}). Clearly these are non-local operations from the point of view of real space, but the important thing is that they also allow us to break the task of connecting two states in the full Fock space into a sequence of operations that act only fixed-momentum subspaces (the minimal number of which defines the complexity). This is, for instance, the way in which cMERA quantum circuits are constructed for free fields \cite{Nozaki:2012zj}, where the procedure of acting with the gates has the nice interpretation of introducing entanglement at different energy scales. Ultimately, the definition of the gates is subjective and we often have to resort to choices of practical convenience, but the underlying idea here is to search for universal features of the complexity that do not depend on this specific choice. We shall see in the following that this is precisely what happens in the present work, namely, the results of \cite{Jefferson:2017sdb} for the free scalar field complexity (obtained using a different set of gates based on real-space intuition) are exactly recovered using our set of gates in the limit of no-quench. 

The target state $\ket{\psi(t)}$ that we want to construct is built from the initial vacuum using essentially the quench Hamiltonian (through time evolution). So, to motivate our choice of gates, it is instructive to write the Hamiltonian associated with our model \eq{scalarfieldaction} in terms of creation and annihilation operators \cite{Alves:2017fjk},\footnote{When $m_\ins=m_\out$ (no quench) this reduces to the standard particle number piece $\sim\int d^{d-1}\kk\,\omega_k\,a_{k}^{\dagger}a_{k}$.}
\begin{equation}\label{eq:Hamiltonian}
H=\frac{1}{2}\int d^{d-1}\kk\,\left[2\big(|\dot{\chi}_{k}|^{2}+\omega_{k}^{2}|\chi_{k}|^{2}\big)a_{k}^{\dagger}a_{k}+\big(\dot{\chi}_{k}^{2}+\omega_{k}^{2}\chi_{k}^{2}\big)a_{k}a_{-k}+\big({\dot{\chi}}_{k}^{*}\,\!^{2}+\omega_{k}^{2}\chi_{k}^{*}\,\!^{2}\big)a_{k}^{\dagger}a_{-k}^{\dagger}\right]+\text{const}\,.
\end{equation}
We see that the Hamiltonian is written as a linear combination 
of operators $\Op_{I,k}$ which mix only $k$ and $-k$ modes (here $I=1,2,3$ or, equivalently, $I=+,-,3$), namely\footnote{Here for convenience we have added an extra piece $\sim a_{-k}^{\,}a_{-k}^\dagger$ in the definition of $\Op_{3,k}$ that did not seem to be present in \eq{Hamiltonian}. Actually it was hidden in the constant term at the end and reduces to first piece after using the commutation relation to convert $a_{-k}^{\,}a_{-k}^\dagger$ into $a_{-k}^\dagger a_{-k}^{\,}$ and renaming the integration variable $k\to-k$.}
\begin{subequations}
\begin{align}
\Op_{+,k}& =\Op_{1,k}+\ii\Op_{2,k} \equiv a_k^\dagger a_{-k}^\dagger \\
\Op_{-,k}& =\Op_{1,k}-\ii\Op_{2,k} \equiv a_k a_{-k} \\
\Op_{3,k}&\equiv \frac{1}{2}(a_k^\dagger a^{\,}_k+ a_{-k}^{\,}a_{-k}^\dagger )\,.
\end{align}
\end{subequations}
It is straightforward to check that (for each $k$) these operators close a SU($1,1$) Lie algebra,\footnote{When writing this we are implicitly assuming that the field is defined is a compact space so that $k$ is a discrete index, otherwise a delta function divergence appears at conciding momenta. There is also a subtlety at the zero modes (when $k=-k$) which we shall ignore here since it plays no role.}
$$[\Op_{3,k},\Op_{\pm,k}]=\pm\Op_{\pm,k},\qquad[\Op_{+,k},\Op_{-,k}]=-2\Op_{3,k}\,,$$
or
$$[\Op_{1,k},\Op_{2,k}]=-\ii\Op_{3,k},\qquad[\Op_{2,k},\Op_{3,k}]=\ii\Op_{1,k},\qquad[\Op_{3,k},\Op_{1,k}]=\ii\Op_{2,k}\,.$$ 
As we shall see below, using them as fundamental gates is enough to build a quantum circuit connecting the target and reference states introduced in the last section. Hence, they constitute a natural set of operators to construct our circuit, and will be our choice. A similar choice of gates has been used recently in \cite{Chapman:2017rqy} to study a different notion of complexity for quantum field theory states. For future convenience we shall adopt the set of gates $\{\Op_{1,k},\Op_{2,k},\Op_{3,k}\}$, so from now on whenever a summation over $I$ appears it refers to $I=1,2,3$.

The quantum circuit is written as
\begin{equation}\label{eq:circuit_def}
U(s) = \mathcal{P}\,e^{\ii\int_0^s\,ds'\,\int\, dk\, Y^I_{k}(s')\Op_{I,k}}
\end{equation}
where the functions $Y^I_k(s)$ define the circuit in the same spirit as in \cite{Jefferson:2017sdb}. We emphasize here the abuse of language when writing $\int dk$, which is to be understood as $\sum_k$ for the field in the compact space or as $V\int dk$ in the continuum limit ($V$ is the spatial volume, which we ignore by taking $V\equiv1$).
Unitarity of $U$ is guaranteed by taking the functions to be real, $\left(Y^I_k\right)^*=Y^I_k$, since the $\Op_{I,k}$ are all Hermitian. Our task is then to find the functions $Y^{I}_{k}\left(s\right)$ that minimize the circuit depth \eq{F_2_distance} while satisfying the boundary conditions
\begin{subequations}\label{eq:bc}
\begin{align}
&U(0)=1\label{eq:bc1}\\
&|\psi\rangle =U(1)|\Omega\rangle\,.\label{eq:bc2}
\end{align}
\end{subequations}

Note that differentiation of \eq{circuit_def} with respect to $s$ yields
\begin{equation}
\int_{-\Lambda}^{\Lambda}dk\,Y^{Ik}\left(s\right)O_{Ik}=\partial_{s}U.U^{-1}\,.
\end{equation}
The main task at this point is to isolate the $Y^{I}_{k}$ to find the line element \eq{line_element}, but this is not possible in the present form. A similar problem was faced by the authors of \cite{Jefferson:2017sdb} when dealing with circuits that build the ground state of free scalar fields. Their trick in that case was to reformulate the problem in matrix language and focus on how the circuit acted on $2\times2$ matrices defining the ground state, which allowed them to solve for the $Y^{I}_{k}$. We shall follow the same steps in the next sub-section.

	\subsection{Matrix representation}

Let us begin by taking a look at how the transformation $U(s)$ acts on the creation and annihilation operators. We first define their transformed versions by
\begin{subequations}
\begin{align}
a^{\,}_k(s) &= U(s)^{-1}\,a^{\,}_k\,U(s)\\
a^\dagger_{-k}(s) &= U(s)^{-1}\,a^{\dagger}_{-k}\,U(s)\,
\end{align}
\end{subequations}
and differentiate with respect to $s$, i.e.,
\begin{equation}
\frac{d}{ds}
\begin{pmatrix}
a_{k}(s)\\
a_{-k}^{\dagger}(s)
\end{pmatrix}
=\frac{dU(s)^{-1}}{ds}\begin{pmatrix}
a_{k}\\
a_{-k}^{\dagger}
\end{pmatrix}
U(s)+U(s)^{-1}\begin{pmatrix}
a_{k}\\
a_{-k}^{\dagger}
\end{pmatrix}\frac{dU(s)}{ds}
\end{equation}
Now we use \eq{circuit_def} and notice that, due to the path-ordering symbol $\mathcal{P}$ (which puts higher values of $s$ always to the left), the derivative of $U$ yields a factor on the left,
\begin{equation}
\frac{dU(s)}{ds}=\left(\ii\int_{-\Lambda}^{\Lambda}dk\,Y^{Ik}(s)O_{Ik}\right)U(s)\,.
\end{equation}
For $U(s)^{-1}$ (defined with anti path-ordering $\bar{\mathcal{P}}$) we have the opposite and the operator appears on the right,
\begin{equation}
\frac{dU(s)^{-1}}{ds}=U(s)^{-1}\left(-\ii\int_{-\Lambda}^{\Lambda}dk\,Y^{Ik}\left(s\right)O_{Ik}\right)\,,
\end{equation}
such that 
\begin{equation}\label{eq:blabla}
\frac{d}{ds}\begin{pmatrix}
a_{k}(s)\\
a_{-k}^{\dagger}(s)
\end{pmatrix}=U(s)^{-1}\left[\left(-\ii\int_{-\Lambda}^{\Lambda}dk'\,Y^{Ik'}(s)O_{Ik'}\right),\begin{pmatrix}
a_{k}\\
a_{-k}^{\dagger}
\end{pmatrix}\right]U(s)\,.
\end{equation}
The commutator is straightforwardly calculated using the canonical commutation relations for $a_k,a^\dagger_k$ and yields a linear combination of $a_k$ and $a^\dagger_k$ themselves. These become $a_k(s)$ and $a^\dagger_k(s)$ after taking into account the $U^{-1}$ and $U$ in equation \eq{blabla}, so that one gets
\begin{equation}\label{eq:vecs}
\frac{d}{ds}
\begin{pmatrix}
a^{\,}_{k}(s)\\
a_{-k}^{\dagger}(s)
\end{pmatrix}
= -G_k(s)
\begin{pmatrix}
a^{\,}_{k}(s)\\
a_{-k}^{\dagger}(s)
\end{pmatrix}
\end{equation}
where $G_k(s)$ is the matrix
\begin{equation}
G_k(s)=
\begin{pmatrix}
-\ii Y^3_{k}(s) &-\ii Y^1_{k}(s)-Y^2_{k}(s)\\
+\ii Y^1_{k}(s)-Y^2_{k}(s) &+\ii Y^3_{k}(s)
\end{pmatrix}\,.
\end{equation}
We immediately recognize $G_{k}\left(s\right)$ as an element of the SU($1,1$) algebra in the $2\times 2$ matrix representation, i.e.,
\begin{equation}
G_k(s) = 2\left[ \ii Y^2_k(s)\Sigma_1 -\ii Y^1_k(s)\Sigma_2 -\ii Y^3_k(s)\Sigma_3\right] \equiv  X^I_k(s)\Sigma_I\,,
\end{equation}
where  
\begin{equation}
\Sigma_1=\frac{1}{2}\begin{pmatrix}0&\ii\\ \,\ii&0\end{pmatrix},\qquad
\Sigma_2=\frac{1}{2}\begin{pmatrix}0&1\\ -1&0\end{pmatrix},\qquad
\Sigma_3=\frac{1}{2}\begin{pmatrix}1&0\\ 0&-1\end{pmatrix}
\end{equation}
satisfy the SU($1,1$) algebra 
$$[\Sigma_1,\Sigma_2]=-\ii\Sigma_3,\qquad [\Sigma_2,\Sigma_3]=\ii\Sigma_1,\qquad [\Sigma_3,\Sigma_1]=\ii\Sigma_2\,.$$
To simplify the notation, in the last step we have introduced
\begin{equation}\label{eq:Xdef}
X^1_k(s)\equiv 2\ii Y^2_k(s),\qquad X^2_k(s)\equiv -2\ii Y^1_k(s),\qquad X^3_k(s)\equiv -2\ii Y^3_k(s)\,.
\end{equation}

Equation \eq{vecs} is formally solved by 
\begin{equation}
\begin{pmatrix}
a^{\,}_{k}(s)\\
a_{-k}^{\dagger}(s)
\end{pmatrix}
= M_k(s)
\begin{pmatrix}
a^{\,}_{k}\\
a_{-k}^{\dagger}
\end{pmatrix}
\end{equation}
with
\begin{equation}\label{eq:Mkdef}
M_k(s)=\mathcal{P}\,e^{-\int_0^s ds'\,G_k(s')}=\mathcal{P}\,e^{-\int_0^s ds'\,X^I_k(s')\Sigma_I}\,.
\end{equation}
The crucial point is that $M_k(s)$, being the exponential of the $\Sigma_I$, is itself nothing but a SU($1,1$) transformation. In other words, $U(s)$ acts on the creation and annihilation operators as 
\begin{equation}\label{eq:UasSU11}
U(s)^{-1}\begin{pmatrix}
a^{\,}_{k}\\
a_{-k}^{\dagger}
\end{pmatrix}U(s)
=M_k(s) \begin{pmatrix}
a^{\,}_{k}\\
a_{-k}^{\dagger}
\end{pmatrix}\,,
\end{equation}
where $M_k(s)$ is a generic SU($1,1$) transformation that we can parametrize as
\begin{equation}\label{eq:Mdef}
M_{k}(s)=
\begin{pmatrix}
{p_k(s)} & {-q_{k}(s)}\\
{-q_{k}^*(s)} & {p_{k}^*(s)}
\end{pmatrix}
\end{equation}
in terms of some $p_{k},q_{k}$ satisfying $|p_{k}|^{2}-|q_{k}|^{2}=1$. Hence our whole circuit can be understood as a sequence of SU($1,1$) transformations $M_k(s)$.

This fact is what allows us to express the functions $Y^I_k(s)$ defining a particular circuit in terms of a particular sequence of SU($1,1$) transformations, solving the problem raised at the end of the previous subsection. Using the fact that the generators $\Sigma_I$ satisfy the orthogonality relation
\begin{equation}\label{eq:SU11orthogonality}
\Tr\left(\Sigma_I\Sigma^\dagger_J\right) = \frac{1}{2}\delta_{IJ}\,,
\end{equation}
equation \eq{Mkdef} can be easily solved for the $X^I_k(s)$ through differentiation with respect to $s$, namely, 
\begin{equation}\label{eq:Xtrace}
X^I_k(s) = 2\,\Tr\left(\partial_sM_k(s)\,M_k(s)^{-1}\,\Sigma^\dagger_I\right)\,.
\end{equation}
This determines the $Y^I_k(s)$ associated with a given SU($1,1$) transformation $M_k(s)$ via \eq{Xdef}.

	\subsection{Boundary conditions for the geodesics}

Having isolated the $Y^I_k(s)$ we now proceed to the second issue raised on the last section, namely, to express the boundary conditions \eq{bc} on $U(s)$ as boundary conditions for the corresponding trajectories in the SU($1,1$) manifold. 

Notice that the condition \eq{constraint_bogoliubov} can be written using \eq{bc2} as 
\begin{equation}
\left(A_k\  B_k\right)\begin{pmatrix}
a^{\,}_{k}\\
a_{-k}^{\dagger}
\end{pmatrix}U(1)|\Omega\rangle = 0\,.
\end{equation}
Applying $U(1)^{-1}$ on both sides and using \eq{UasSU11} we have 
\begin{equation}
\left(A_k\  B_k\right)M_k(1)\begin{pmatrix}
a^{\,}_{k}\\
a_{-k}^{\dagger}
\end{pmatrix}|\Omega\rangle = 0\,,
\end{equation}
and by comparing with the defining constraint \eq{reftargetdef1} for the reference state $|\Omega(t)\rangle$ we immediately identify
\begin{equation}
\left(A_k\  B_k\right)M_k(1) = N_k \left(\alpha_k\  \beta_k\right)\,,
\end{equation}
where $N_k$ is a proportionality constant.

This is a pair of equations that can be solved for $p_k(1)$ and $q_k(1)$ (the matrix components of $M_k(1)$) in terms of $A_k,B_k,\alpha_k,\beta_k$ (the coefficients characterizing the reference and target states). The boundary conditions \eq{bc} are then rephrased as
\begin{subequations}\label{eq:bcpq}
\begin{align}
{p_{k}^*(1)}&=N_{k}^{*}\alpha_{k}^{*}A_{k}-N_k\beta_{k}B_{k}^{*}\\
{q_{k}(1)}&=N_{k}^{*}\alpha_{k}^{*}B_{k}-N_k\beta_{k}A_{k}^{*}\\
{p_{k}^*(0)}&=1\\
{q_{k}(0)}&=0\,,
\end{align}
\end{subequations}
where we included also the last two which correspond simply to the statement that $M_k(0)=1$. The proportionality constant is fixed by requiring compatibility with the SU($1,1$) normalization $|p_k(1)|^2-|q_k(1)|^2=1$, which leads to 
\begin{equation}
|N_{k}|=\frac{1}{\sqrt{|\alpha_{k}|^{2}-|\beta_{k}|^{2}}}\,.
\end{equation}
We see that there is still a phase factor freedom in $N_{k}$. This is related to an ambiguity in the definition of our reference and target states which we shall discuss in detail in the next section, so for now one can simply take $N_{k}$ to be real without loss of generality and postpone the discussion of the phases to the next section. 

	\subsection{Phase factor ambiguity}

It is important to notice (see \cite{Nozaki:2012zj} for a similar discussion  in the context of cMERA) that there is an ambiguity in the definitions \eq{reftargetdef} of our reference and target states. If one multiplies any of the constraints by a time-dependent phase factor we still get the same physical constraint, i.e.,
\begin{subequations}
\begin{align}
e^{\ii\eta_k(t)}\left(\alpha_{k}\,a_{k}+\beta_{k}\,a_{-k}^{\dagger}\right)|\Omega\rangle&=0\notag\\
e^{\ii\tilde\eta_k(t)}\left(A_{k}\,a_{k}+B_{k}\,a_{-k}^{\dagger}\right)|\psi\rangle&=0\,.\notag
\end{align}
\end{subequations}
Equivalently, this means that there is the ambiguity of replacing $(\alpha_k,\beta_k)\to e^{\ii\eta_k}(\alpha_k,\beta_k)$ and $(A_k,B_k)\to e^{\ii\tilde\eta_k}(A_k,B_k)$.\footnote{One could in principle consider including also a real number $r_k(t)$ in front of the phase factors. However, for the case of $A_k,B_k$ the condition $|A_{k}|^{2}-|B_{k}|^{2}=1$ forces $r_k=1$ while in the case of $\alpha_k,\beta_k$ this number gets cancelled from the boundary conditions \eq{bcpq} by the normalization factor and hence plays no role.} Since they appear in a mixed way in \eq{bcpq}, there is a resulting relative phase ambiguity between the two terms that is nonvanishing. Even though the initial and final states are obviously not affected by this choice of phase, the circuit connecting them in general is and, as a result, the circuit depth associated to a given path can depend on it. In practice, this means that the boundary conditions \eq{bcpq} for $p_k,q_k$ have to be replaced by (here $\xi_k\equiv\tilde\eta_k-\eta_k$)
\begin{subequations}\label{eq:bcpqmodified}
\begin{align}
{p_{k}^*(1)}&=\frac{e^{\ii\xi_k}\alpha_{k}^{*}\,A_{k}-e^{-\ii\xi_k}\beta_{k}\,B_{k}^{*}}{\sqrt{|\alpha_{k}|^{2}-|\beta_{k}|^{2}}}\label{eq:bcpqmodified1}\\
{q_{k}(1)}&=\frac{e^{\ii\xi_k}\alpha_{k}^{*}\,B_{k}-e^{-\ii\xi_k}\beta_{k}\,A_{k}^{*}}{\sqrt{|\alpha_{k}|^{2}-|\beta_{k}|^{2}}}\label{eq:bcpqmodified2}\\
{p_{k}^*(0)}&=1\\
{q_{k}(0)}&=0\,,
\end{align}
\end{subequations}
and, when seeking for the optimal circuit to find the complexity, in addition to finding the geodesics associated with the metric \eq{metrick} we also need to find the choice of relative phase $\xi_k(t)$ that minimizes the circuit depth. We will show explicitly that such a choice of $\xi_k$ is \emph{not} the simplest one ($\xi_k=0$).

\section{Time evolution of the complexity}\label{time_evolution_if_the_cpxt}
	\subsection{Geometrizing the problem}

We are now ready to geometrize and solve the complexity problem. 
The first step is to parametrize the manifold of SU($1,1$) matrices using the set of coordinates $\{\theta_k,\phi_k,\chi_k\}$ defined by
\begin{equation}\label{eq:coordsdef}
{p_k^*(s)}\equiv \cosh\theta_{k}(s)\,e^{\ii\phi_{k}(s)}\qquad\qquad 
{q_k(s)}\equiv \sinh\theta_{k}(s)\,e^{-\ii\chi_{k}(s)}\,,
\end{equation}
which clearly satisfy $|p_k(s)|^2-|q_k(s)|^2=1$. The boundary conditions \eq{bcpqmodified} then can be translated as
\begin{subequations}\label{eq:bccoords}
\begin{align}
\phi_{k}(0)&=0\\
\theta_{k}(0)&=0\\
\chi_{k}(0)&=\chi_{0}\\
\phi_{k}(1)&=\text{Arg}\left({p_{k}^*(1)}\right)\\
\theta_{k}(1)&=\text{arccosh}\left(|{p_{k}^*(1)}|\right)=\text{arcsinh}\left(|{q_{k}(1)}|\right)\\
\chi_{k}(1)&=-\text{Arg}\left({q_{k}(1)}\right)\,
\end{align}
\end{subequations}
with $\chi_{0}$ an arbitrary constant. Here, $p_k^*(1)$ and $q_k(1)$ are to be understood as shorthand for the expressions \eq{bcpqmodified1},\eq{bcpqmodified2}.
 
The functions $Y^I_k(s)$ characterizing our circuit can be expressed in terms of the coordinates using equations \eq{Xdef}, \eq{Mdef}, \eq{Xtrace} and \eq{coordsdef}. One finds
\begin{subequations}
\begin{align}
Y^1_{k}&=-\sin(\phi_{k}-\chi_{k})\,\theta'_{k}+\frac{1}{2}\cos(\phi_{k}-\chi_{k})\sinh(2\theta_{k})\left(\phi'_{k}+\chi'_{k}\right)\\
Y^2_{k}&=-\cos(\phi_{k}-\chi_{k})\,\theta'_{k}-\frac{1}{2}\sin(\phi_{k}-\chi_{k})\sinh(2\theta_{k})\left(\phi'_{k}+\chi'_{k}\right)\\
Y^3_{k}&=-\cosh^2(\theta_{k})\,\phi'_{k}-\sinh^2(\theta_{k})\,\chi'_{k}\,,
\end{align}
\end{subequations}
where a prime is used here to denote $\frac{d}{ds}$.

The circuit depth \eq{F_2_distance} associated with a given curve $Y^I_k(s)\equiv\{\theta_k(s),\phi_k(s),\chi_k(s)\}$ in the SU($1,1$) manifold is then given by 
\begin{align}\label{eq:CD}
\mathcal{D}_{F_2}(U) &= 
\int_0^1 ds\,\sqrt{\int_{-\Lambda}^\Lambda dk\,\big(\,|Y^1_k(s)|^2+|Y^2_k(s)|^2+|Y^3_k(s)|^2\big)}\notag\\
&= \int_0^1 ds\,\sqrt{\int_{-\Lambda}^\Lambda dk\,\left[
{\theta'_{k}}^2+
\frac{1}{4}{v'_{k}}^{2}+
\frac{1}{4}\cosh(4\theta_{k})\,{u'_{k}}^{2}+
\frac{1}{2}\cosh(2\theta_{k})\,u'_{k}\,v'_{k}\right]}\,,
\end{align}
where we defined the new variables
\begin{subequations}\label{eq:uvdef}
\begin{align}
u_{k}&=\phi_{k}+\chi_{k}\\
v_{k}&=\phi_{k}-\chi_{k}\,.
\end{align}
\end{subequations}
The problem now becomes a geometrical one by noticing that this can be seen as the length of a particle's trajectory in a curved Riemannian manifold with metric 
\begin{equation}
ds^{2}=\int_{-\Lambda}^{\Lambda}dk\left[{d\theta_{k}}^{2}+\frac{1}{4}dv_{k}^{2}+\frac{1}{4}\cosh(4\theta_{k})\,du_{k}^{2}+\frac{1}{2}\cosh(2\theta_{k})\,du_{k}\,dv_{k}\right]\,
\end{equation}
(again, we stress that there is an abuse of notation here: the way to make sense of this is to formulate the problem in a compact space so that $\int dk$ is replaced by a discrete sum over $k$).
The optimal path is simply a geodesic in this geometry, the length of which defines the circuit complexity.

One can see that the metric is block diagonal with respect to the momentum-mode label $k$, i.e., $ds^{2}=\int_{-\Lambda}^{\Lambda}dk\,ds_{k}^{2}$. 
It is also important to notice that each block $k$ depends only on the coordinates $\{\theta_{k},\psi_{k},\phi_{k}\}$ associated with that same momentum $k$ (there is no mode mixing). This means that we can find the curves minimizing the length in each of these momentum subspaces and add them together to find the geodesic in the whole manifold, which can also be checked by simply looking at the geodesic equation. 
Therefore the main goal now is to calculate the geodesic associated with the metric
\begin{equation}\label{eq:metrick}
ds_k^2={d\theta_{k}}^{2}+\frac{1}{4}dv_{k}^{2}+\frac{1}{4}\cosh(4\theta_{k})\,du_{k}^{2}+\frac{1}{2}\cosh(2\theta_{k})\,du_{k}\,dv_{k}
\end{equation}
subject to the boundary conditions \eq{bccoords}. 

Before delving into the calculations to find the geodesics, a comment is in order here. The metric above is diffeomorphic to the one found in \cite{Jefferson:2017sdb} restricted to a hypersurface $y=\text{const}$. This should not come as a surprise since our choice of gates $\{\Op_{1,k},\Op_{2,k},\Op_{3,k}\}$ leads to geodesics in the space of SU($1,1$) transformations, while the gates used by the authors of \cite{Jefferson:2017sdb} lead to geodesics in the space of GL($2,\mathbb{R}$) transformations. It is well-known that SU($1,1$) is isomorphic to SL($2,\mathbb{R}$), a subgroup of GL($2,\mathbb{R}$), so at the end of the day we should expect to deal with a submanifold of the one appearing in \cite{Jefferson:2017sdb}.

	\subsection{Symmetries and Killing vectors}

The task of finding the geodesics associated with the metric \eq{metrick} can be simplified by taking advantage of symmetries and their corresponding conserved quantities. By inspection one can check that the metric \eq{metrick} possesses a trivial Killing vector, $k_0\equiv\partial_{v_k}$, but the question remains whether there are others. As shown in \cite{Jefferson:2017sdb}, to each generator of the symmetry group there must be a corresponding Killing vector. This is because of the way the metric is defined, via the functions $Y^I_k(s)$ in \eq{Xtrace}: if we redefine $M_{k}(s)\rightarrow M_{k}(s)\tilde{M}$, where $\tilde{M}$ is a constant SU($1,1$) matrix, the $Y^I_k$ (and consequently the metric) remain unchanged. So in our case we should expect at least $3$ Killing vectors (one for each SU($1,1$) generator). It turns out that none of them happens to be the $k_0$ above, so there are actually $4$ in total.\footnote{$k_0$ can be seen as an \lq\lq accidental\rq\rq symmetry reminiscent of our choice of $G_{IJ}=\delta_{IJ}$ as the $F_2$-metric in \eq{F_2_distance} \cite{Jefferson:2017sdb}.}

Following \cite{Jefferson:2017sdb}, let us take $\tilde{M}=e^{\epsilon_{I}\Sigma^{I}}$ with $\epsilon_{I}$ ($I=1,2,3$) constants. To first order in $\epsilon_I$, the statement is that the metric is invariant under
\begin{equation}
\delta M_{k}=M_k\,\Sigma^{I}\epsilon_{I}\,.
\end{equation}
To find the corresponding Killing vectors, we assume this change in $M_k$ is coming from a diffeomorphism $\delta x^j$, i.e,
\begin{equation}
\delta M_{k}=\partial_{j}M_{k}\,\delta x^{j}\,,
\end{equation}
and use \eq{SU11orthogonality} to find 
\begin{equation}
\epsilon_{I}=2\,\Tr\left(M_k^{-1}\,\partial_{j}M_{k}\,\Sigma_{I}^{\dagger}\right)\delta x^{j}\,.
\end{equation}
This can be inverted in the ($j,I$) indices to yield
\begin{equation}
\delta x_j=(k_I)_j\,\epsilon^{I}\,
\end{equation}
with the Killing vectors $(k_I)_j$ given by
\begin{equation}\label{eq:Killingdef}
(k_I)_j=\frac{1}{2}\,\left[\Tr\left(M_k^{-1}\,\partial_{j}M_{k}\,\Sigma_{I}^{\dagger}\right)\right]^{-1}\,.
\end{equation}
Indeed, we see that there are as many of them as there are SU($1,1$) generators $\Sigma_I$.

In terms of the coordinates $\{\theta_k(s),\phi_k(s),\chi_k(s)\}$, the Killing vectors above read
\begin{subequations}
\begin{align}
k_{1}&=\frac{\ii}{2}\left[\cos(u_{k})\partial_{\theta_{k}}-\sin(u_{k})\left(\tanh(\theta_{k})+\coth(\theta_{k})\right)\partial_{u}-\sin(u_{k})\left(\tanh(\theta_{k})-\coth(\theta_{k})\right)\partial_{v_{k}}\right]\\
k_{2}&=\frac{\ii}{2}\left[\sin(u_{k})\partial_{\theta_{k}}+\cos(u_{k})\left(\tanh(\theta_{k})+\coth(\theta_{k})\right)\partial_{u}+\cos(u_{k})\left(\tanh(\theta_{k})-\coth(\theta_{k})\right)\partial_{v_{k}}\right]\\
k_{3}&=-\ii\partial_{\theta_k}\,,
\end{align}
\end{subequations}
which together with $k_0=\partial_{v_k}$ discussed before consist of all the isometries of the metric \eq{metrick}.

To each of these Killing vectors one can associate conserved momenta $c_{I}=\left(k_{I}\right)^{i}g_{ij}{x'}^{j}$ which remain constant along the geodesics and can be used to simplify the task of integrating the geodesic equations. Their explicit expressions (apart from overall numerical factors) are the following
\begin{subequations}\label{eq:conserved}
\begin{align}
c_{0} &= {v'_{k}}+\cosh(2\theta_{k})\,{u'_{k}}\label{eq:conserved1}\\
c_{1} 
&= \cos(u_k)\,{\theta'_k} -\frac{1}{2}\sinh(4\theta_k)\,\sin(u_k)\,u'_k - \frac{1}{2}\sinh(2\theta_k)\sin (u_k)\,v'_k\label{eq:conserved2}\\
c_2 &= \sin(u_k)\,{\theta'_k} + \frac{1}{2}\sinh(4\theta_k)\,\cos(u_k)\,u'_k + \frac{1}{2}\sinh(2\theta_k)\cos(u_k)\,v'_k\label{eq:conserved3}\\
c_3 &= \cosh(4\theta_k)\,u'_k + \cosh(2\theta_k)\,v'_k \,.\label{eq:conserved4}
\end{align}
\end{subequations}

We now proceed to use these equations to find the geodesics.

	\subsection{Geodesics and the complexity}

From \eq{conserved1} and \eq{conserved4} it follows that
\begin{equation}
u'_{k}=\frac{c_{3}-c_{0}\cosh(2\theta_{k})}{\sinh^{2}(2\theta_{k})}\qquad\text{and}\qquad v'_{k}=\frac{c_0\cosh(4\theta_k)-c_{3}\cosh(2\theta_{k})}{\sinh^{2}(2\theta_{k})}\,.
\end{equation}
Now, the boundary conditions at $s=0$ (see \eq{bccoords}) demand that $c_0=c_3=0$, otherwise $u'_k,v'_k$ (and hence the length of the curve) would be divergent at $s=0$. This implies $u'_k=0,v'_k=0$, i.e.,
\begin{subequations}\label{eq:uvsol}
\begin{align}
u_{k}(s) &=u_k(0) = \chi_0\\
v_{k}(s) &= v_k(0) = -\chi_0\,,
\end{align}
\end{subequations}
where in the last step we have expressed the result in terms of the boundary condition for $\chi_k$ using the definition \eq{uvdef} of $u_k,v_k$ in terms of the original variables $\phi_k,\chi_k$. By plugging these results into \eq{conserved2} and \eq{conserved3} and imposing the boundary condition $\theta_k(0)=0$ we get simply
\begin{equation}\label{eq:thetapartialsol}
\theta_k(s) = \frac{c_1}{\cos(\chi_0)}\,s\qquad\text{and}\qquad \frac{c_2}{c_1}=\tan(\chi_0)\,.
\end{equation}

What remains is to determine the constant $c_1$ in addition to fixing the global phase factor $\xi_k$ using the boundary conditions at $s=1$. These can be rewritten (using \eq{uvsol} and \eq{thetapartialsol}) as
\begin{subequations}\label{eq:bcalmostdone}
\begin{align}
\phi_{k}(1)&=\text{Arg}\left({p_{k}^*(1)}\right) = \arctan\left(\frac{\text{Im}\,{p_k^*(1)}}{\text{Re}\,{p_k^*(1)}}\right) = 0 \label{eq:bcalmostdone1}\\ 
\theta_{k}(1)&=\text{arccosh}\left(|{p_{k}^*(1)}|\right) = \text{arccosh}\left(\sqrt{\left(\text{Re}\,{p_k^*(1)}\right)^2+\left(\text{Im}\,{p_k^*(1)}\right)^2}\right) = \frac{c_1}{\cos(\chi_0)}\label{eq:bcalmostdone2}\\
\chi_{k}(1)&=-\text{Arg}\left({q_{k}(1)}\right) = -\arctan\left(\frac{\text{Im}\,{q_k(1)}}{\text{Re}\,{q_k(1)}}\right) = \chi_0\label{eq:bcalmostdone3}
\end{align}
\end{subequations}
with $p_k(1),q_k(1)$ given in \eq{bcpqmodified}. 
Equation \eq{bcalmostdone1} allows us to fix the phase ambiguity, namely we have to solve 
\begin{equation}
\text{Im}\,{p_k^*(1)} = \frac{1}{\sqrt{|\alpha_k|^2-|\beta_k|^2}}\text{Im}\left(\alpha_{k}^{*}\,e^{\ii\xi_k}A_{k}-\beta_{k}\,e^{-\ii\xi_k}B_{k}^{*}\right) =0\,
\end{equation}
for $\xi_k$. The solution is easily found to be
\begin{equation}
\xi_k = \arctan\left(\frac{\text{Im}(\beta_k\,B_k^*-\alpha_k^*\,A_k)}{\text{Re}(\beta_k\,B_k^*+\alpha_k^*\,A_k)}\right),
\end{equation}
which then implies
\begin{equation}\label{eq:Rep}
\text{Re}\,{p_k^*(1)} = \frac{1}{\sqrt{|\alpha_k|^2-|\beta_k|^2}}\,\frac{\left|\alpha_k\right|^2\,\left|A_k\right|^2-\left|\beta_k\right|^2\,\left|B_k\right|^2}{\sqrt{\left[\text{Re}(\alpha_k^*\,A_k+\beta_k\,B_k^*)\right]^2+\left[\text{Im}(\alpha_k^*\,A_k-\beta_k\,B_k^*)\right]^2}}\,.
\end{equation}
This completely fixes $\theta_k(1)$ in \eq{bcalmostdone2} in terms of the target and reference state data, i.e.,
\begin{equation}\label{eq:theta1Rep}
\theta_{k}(1)=\text{arccosh}\left(|\text{Re}\,{p_{k}^*(1)}|\right)\,.
\end{equation}
The remaining constants $\chi_0,c_1$ are also uniquely determined in terms of these data by equations \eq{bcalmostdone2} and \eq{bcalmostdone3}, but the expressions will not be needed here.

Having obtained the geodesic $\{\theta_k(s)=\theta_k(1)\,s,u_k(s)=\chi_0,v_k(s)=-\chi_0\}$, the complexity is then readily found by substituting it into expression \eq{CD} to find its length (the minimal circuit depth). Since $u_k$ and $v_k$ have vanishing derivatives, we see that the complexity is fully determined by the boundary value of $\theta_k(s)$, namely, 
\begin{align}\label{eq:CDfinal}
C_{F_2}(U)&=\text{min}\left(\mathcal{D}_{F_2}(U)\right)= \sqrt{\int_{-\Lambda}^\Lambda dk\,\theta_k(1)^2}\,.
\end{align}
Using \eq{theta1Rep} and \eq{Rep} one can write this explicitly as
\begin{equation}\label{eq:cpxt_finalanswer}
C_{F_2}(U) = \sqrt{\int_{-\Lambda}^\Lambda dk\,\text{arccosh}^2\left(\left|\frac{1}{\sqrt{|\alpha_k|^2-|\beta_k|^2}}\,\frac{\left|\alpha_k\right|^2\,\left|A_k\right|^2-\left|\beta_k\right|^2\,\left|B_k\right|^2}{\sqrt{\left[\text{Re}(\alpha_k^*\,A_k+\beta_k\,B_k^*)\right]^2+\left[\text{Im}(\alpha_k^*\,A_k-\beta_k\,B_k^*)\right]^2}}\right|\right)}\,.
\end{equation}
This analytic expression for the complexity in terms of the reference and target state data ($A_k,B_k,\alpha_k(t),\beta_k(t)$) given in \eq{bogoliubovcoeffs} and \eq{alphabetadef}, with the out-modes given in \eq{out_modes} is the main result of the present paper. 

The time evolution due to the quench is encoded in the coefficients $\alpha_k(t),\beta_k(t)$ and is highly nontrivial, via hypergeometric functions. For this reason the integral over $k$ (or a discrete summation over $k$, to be more precise) cannot be carried out analytically, but it is straightforward to evaluate it numerically for specific choices of the parameters characterizing the quench if we impose a finite cutoff $\Lambda$. We shall do that in a moment, but first let us analyze some interesting limits.

		\subsubsection{Static limit: the vacuum complexity}

As a consistency check, we can particularize to the static or no quench case that corresponds to $m_\text{out}=m_\text{in}$. Equivalently, this can be seen as the early time limit $t\to-\infty$ of the time evolution following the mass quench. In this limit, the construction above gives simply the ground state complexity of a massive scalar field (with mass $m_\ins$) previously studied in \cite{Jefferson:2017sdb} (remember that our choices of reference and target states in this limit become exactly the same as those of \cite{Jefferson:2017sdb}). 

The coefficients $A_k,B_k,\alpha_k,\beta_k$ take the very simple form
\begin{equation}
A_k=1\,,\qquad B_k=0\,,\qquad \alpha_k(t)=\frac{M+\omega_\ins}{2\sqrt{2M\omega_\ins}}\,e^{-\ii\omega_\ins t},\qquad \beta_k(t)=\frac{M-\omega_\ins}{2\sqrt{2M\omega_\ins}}\,e^{\ii\omega_\ins t}\,,
\end{equation}
and the complexity \eq{cpxt_finalanswer} reduces to
\begin{equation}\label{eq:cpxtnoquencharcosh}
C_{F_2}(U) = \sqrt{\int_{-\Lambda}^\Lambda dk\,\text{arccosh}^2\left(\frac{M+\omega_\ins}{\sqrt{4M\omega_\ins}}\right)} \,.
\end{equation}
Using the identity 
\begin{equation}
\text{arccosh}(x)=\log\left(x+\sqrt{x^{2}-1}\right)
\end{equation}
we get
\begin{equation}\label{eq:cpxtvacuum}
C_{F_2}^\text{vacuum} = \frac{1}{2}\,\sqrt{\int_{-\Lambda}^\Lambda dk\,\left(\log\frac{\omega_\ins}{M}\right)^{2}}\,,
\end{equation}
which is exactly (the continuum version of) the expression for the complexity in a discretized massive scalar field computed in \cite{Jefferson:2017sdb}. It is interesting to stress that the choice of elementary gates used in \cite{Jefferson:2017sdb} differs significantly from the one used here, yet both results for the complexities agree. This indicates some degree of universality of the complexity with respect to the choice of elementary gates. 

		\subsubsection{Sudden quench limit}\label{sec:fast}

Another interesting limit to look at is the infinitely fast or sudden quench limit $\delta t \rightarrow 0$, where the mass profile \eq{massquenchtanh} reduces to a step function. For $t<0$ there is essentially no dynamics since $m=m_\ins$ is held fixed and we get simply the vacuum result above $C_{F_2}^\text{vacuum}$ for the complexity. For $t>0$, on the other hand, we have
\begin{equation}
A_k=\frac{\omega_+}{\sqrt{\omega_\text{in}\omega_\text{out}}},\qquad B_k=-\frac{\omega_-}{\sqrt{\omega_\text{in}\omega_\text{out}}},\qquad 
\alpha_k(t)=\frac{M+\omega_\out}{2\sqrt{2M\omega_\text{out}}}\,e^{-\ii\omega_\out t}, \qquad
\beta_k(t)=\frac{M-\omega_\out}{2\sqrt{2M\omega_\text{out}}}\,e^{\ii\omega_\out t}\,,
\end{equation}
and the complexity \eq{cpxt_finalanswer} reduces to
\begin{equation}\label{eq:delta_t_zero}
C_{F_2}=\sqrt{\int_{-\Lambda}^\Lambda dk\,\text{arccosh}^2\left( 
\frac{|(M+\omega_\out)^2\omega_+^2-(M-\omega_\out)^2\omega_-^2|/\sqrt{4M\omega_\ins\,\omega_\out^2}}{\sqrt{(M+\omega_\out)^2\omega_+^2+(M-\omega_\out)^2\omega_-^2-2(M^2-\omega_\out^2)\omega_+\omega_-(\cos\omega_\out t-\sin\omega_\out t)}}
\right)}\,.
\end{equation}
In other words, in this limit the complexity jumps from the constant value \eq{cpxtvacuum} at $t<0$ to an oscillatory behavior at $t>0$ that persists forever. The frequency of oscillation is not easy to guess though, since $\omega_\out$ depends on the momentum variable $k$ which is being integrated over. As we shall see below, this behavior is compatible with the smooth quench results, where the same late time oscillations appear for $t\gtrsim\delta t$.

		\subsubsection{Small and large reference scale $M$}

One can also play with the free scale $M$ that characterizes the reference state and obtain reasonably simple exact expressions for the complexity \eq{cpxt_finalanswer} in the extreme limits where $M\to0$ and $M\to\infty$. Since $M$ appears only in the functions $\alpha_k(t),\beta_k(t)$, taking these limits is straightforward using \eq{alphabetadef}, namely
\begin{subequations}\label{eq:Mlimits}
\begin{align}
M\to\infty:&\qquad \alpha_k(t) = \frac{\sqrt{M}}{2}\chi_k(t)\,,\quad \beta_k(t) = \frac{\sqrt{M}}{2}\chi^*_{-k}(t)\\
M\to0:     &\qquad \alpha_k(t) = \frac{\ii}{2\sqrt{M}}\dot\chi_k(t)\,,\quad \beta_k(t) = \frac{\ii}{2\sqrt{M}}\dot\chi^*_{-k}(t)\,
\end{align}
\end{subequations}
while the difference $|\alpha_k|^2-|\beta_k|^2 = \text{Im}\left(\chi_k\,\dot\chi_k^*\right)$ is independent of $M$. This leads to 
\begin{subequations}
\begin{align}
M\to\infty:&\quad 
C_{F_2}(U) = \sqrt{\int_{-\Lambda}^\Lambda dk\,\log^2\left(\frac{\sqrt{M}}{\sqrt{\text{Im}\left(\chi_k\,\dot\chi_k^*\right)}}\,\frac{|\chi_k|^2}{\sqrt{\left[\text{Re}(\chi_k^*\,(A_k+B_k^*)\right]^2+\left[\text{Im}(\chi_k^*\,(A_k-B_k^*))\right]^2}}\right)}\\
M\to0:     &\quad 
C_{F_2}(U) = \sqrt{\int_{-\Lambda}^\Lambda dk\,\log^2\left(\frac{1}{\sqrt{M\,\text{Im}\left(\chi_k\,\dot\chi_k^*\right)}}\,\frac{|\dot\chi_k|^2}{\sqrt{\left[\text{Re}(\dot\chi_k^*\,(A_k+B_k^*)\right]^2+\left[\text{Im}(\dot\chi_k^*\,(A_k-B_k^*))\right]^2}}\right)}
\end{align}
\end{subequations}
where we have used $\text{arccosh}(x)\approx\log(2x)$ ($x\to\infty$) to convert the integrand into a logarithm. The main difference between the two limits appears in the second ratio inside the argument of the log, being the time dependence dictated by $\dot\chi_k(t)$ as $M\to0$ and by $\chi_k(t)$ as $M\to\infty$ as already anticipated by \eq{Mlimits}. Intermediate scales of $M$ are therefore characterized by an interplay between these two behaviors.

		\subsubsection{UV behavior}

It is important to notice that the ground state complexity \eq{cpxt_finalanswer} is UV divergent. This is easily seen by analyzing the large $k$ behavior of the integrand, namely
\begin{equation}
C_{F_2} = \sqrt{\int_{-\Lambda}^\Lambda dk\,\text{arccosh}^2\left(\sqrt{\frac{|k|}{4M}}\right)+\cdots} \sim \Lambda\left(\log\frac{\Lambda}{M}\right)^2+\cdots\,.
\end{equation}
However, the leading UV divergence happens to be exactly the same as the one of the ground state complexity \eq{cpxtnoquencharcosh}, since $\omega_\ins\sim |k|$ for large $k$. It is tempting then to subtract this constant divergent piece and introduce the \emph{\lq\lq renormalized\rq\rq~complexity} (i.e., the relative complexity of the the instantaneous state $\ket{\psi(t)}$ with respect to the ground state one)
\begin{equation}\label{eq:cpxtrenorm}
\Delta C_{F_2}(t) \equiv  C_{F_2}(t)- C_{F_2}^\text{vacuum}\,,
\end{equation}
which is guaranteed to be UV finite during the whole time evolution. By definition, this quantity vanishes at early times since the pre-quench state is nothing but the vacuum of the initial Hamiltonian. In the following we shall study it numerically for different choices of quench parameters as well as reference state data.

		\subsubsection{Numerical analysis}
        
Having looked at special solvable limits of interest, we now analyze the renormalized complexity $\Delta C_{F_2}(t)$ introduced in \eq{cpxtrenorm} in its full generality using numerical methods. As discussed above, this object is UV finite unlike the complexity \eq{cpxt_finalanswer} itself. In practice, the analysis shall be carried for a fixed and finite (but large) value for the momentum cutoff $\Lambda$. We emphasize once again that the integrand appearing in \eq{cpxt_finalanswer} is known exactly in terms of the reference and target state data, but its $k$-dependence is highly nontrivial such that doing the integration analytically becomes hopeless. That is why we need to resort to numerical integration, at this very last step of the analysis.

\begin{figure}[ht]
    \centering
    \subfigure[Mass-increasing quenches with fixed $\delta t,M$.]{\includegraphics[height=2in]{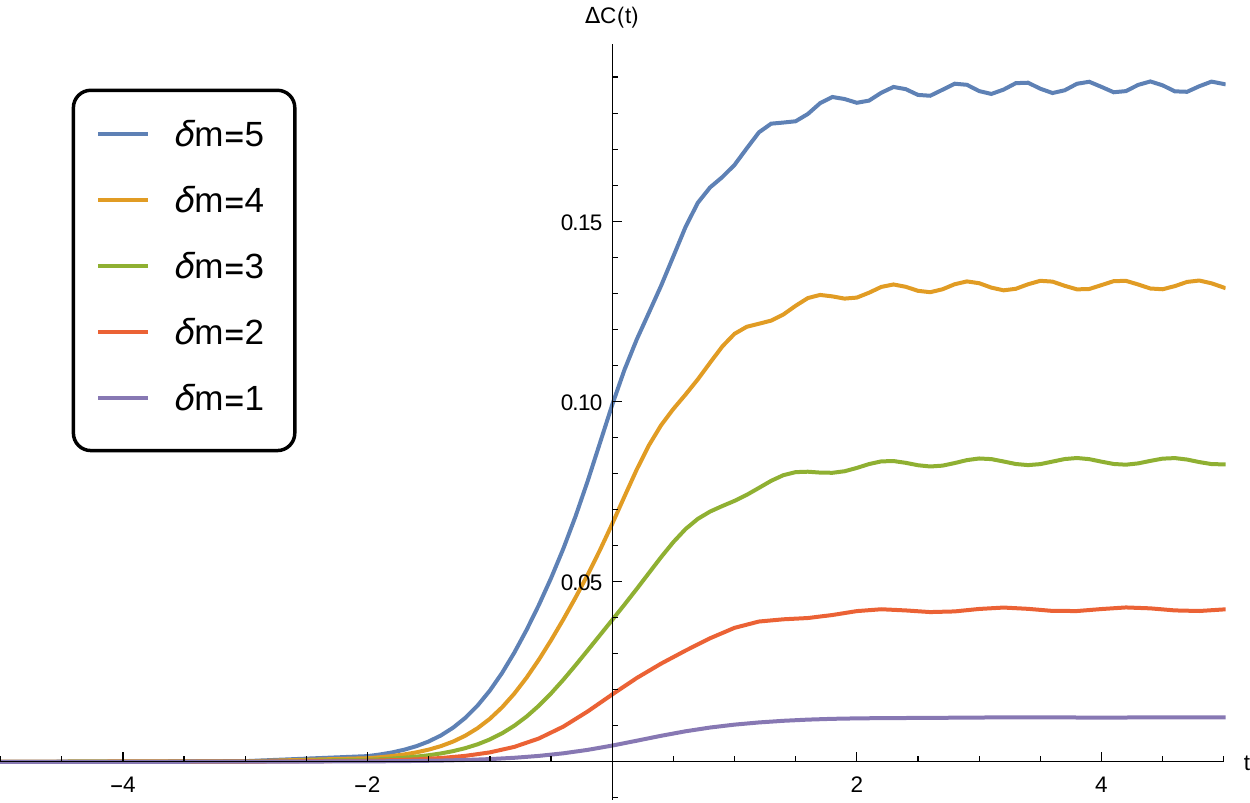}} \qquad
    \subfigure[Mass-decreasing quenches with fixed $\delta t,M$.]{\includegraphics[height=2in]{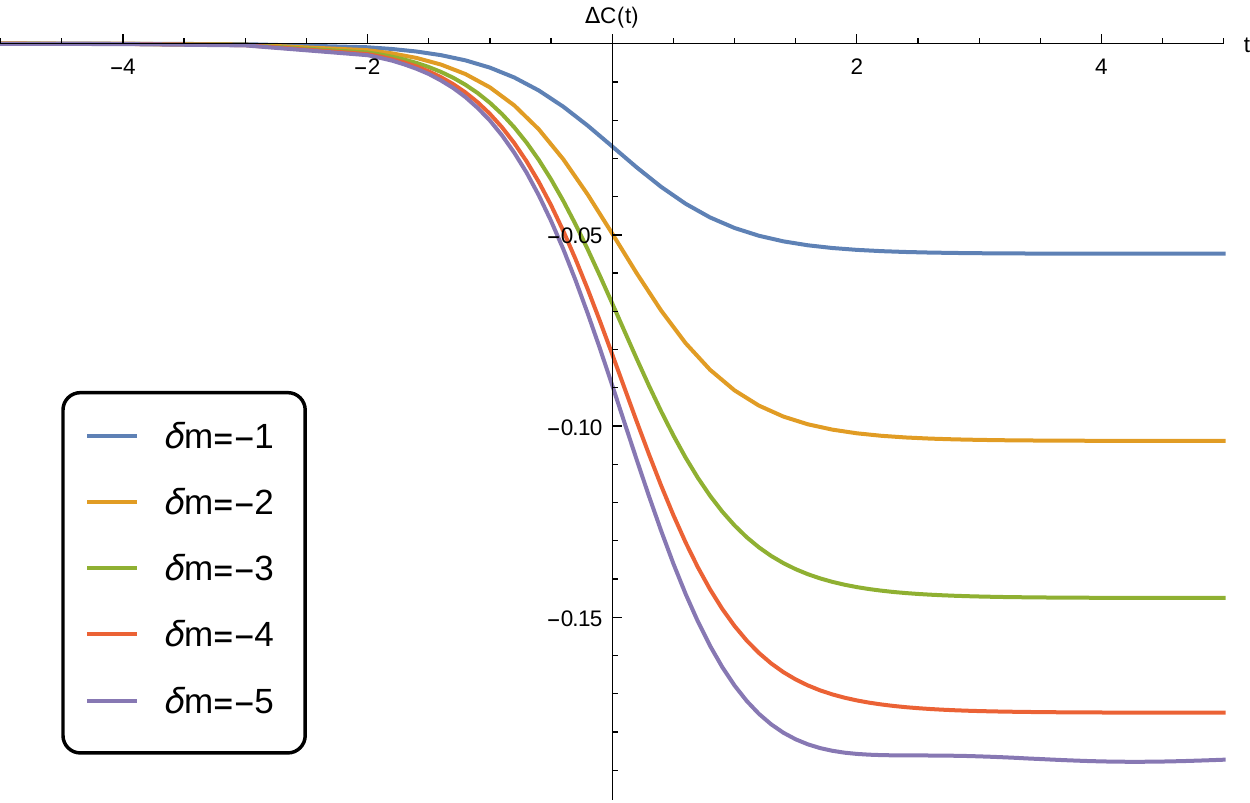}}\\
    \subfigure[Different quench rates with fixed $\delta m,M$.]{\includegraphics[height=2in]{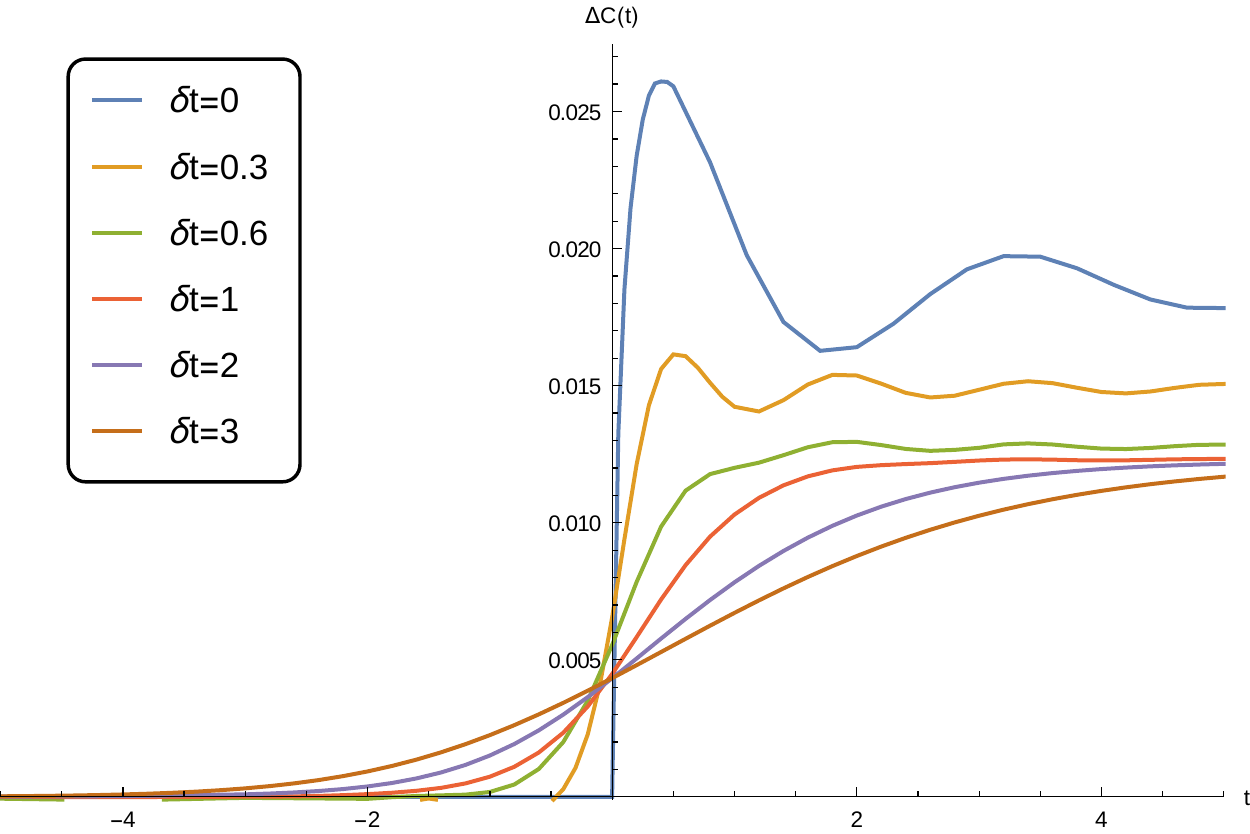}}\qquad
    \subfigure[Different reference scales with fixed $\delta t,\delta m$.]{\includegraphics[height=2in]{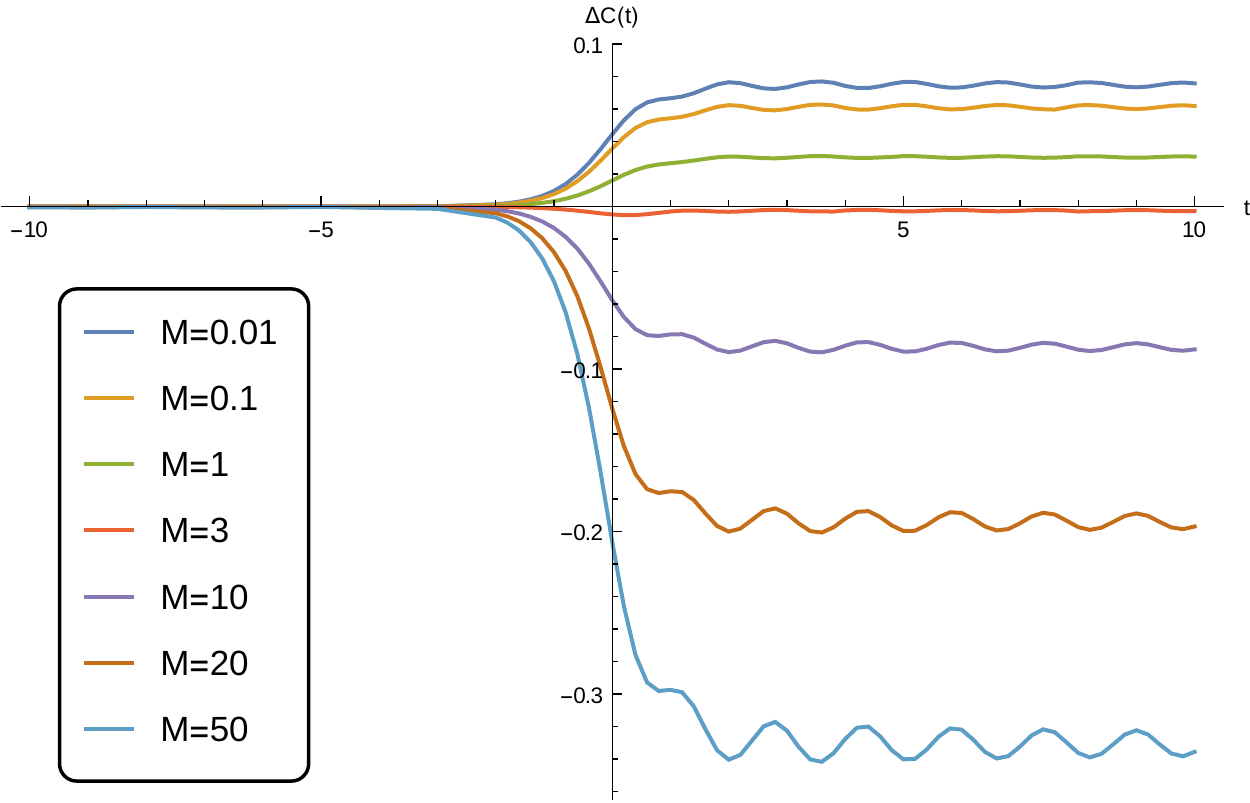}}
	\caption{Time dependence of the complexity for many combinations of the mass difference $\delta m=m_\out-m_\ins$, the quench rate $\delta t$, and the scale $M$ characterizing the reference state. All the plots were made with specific numerical values of the parameters chosen for presentation purposes, but the qualitative behavior above was checked to hold generically. Figure (a) has $\delta t=1,M=2,m_\ins=1$ and $m_\out$ is changed as indicated by the inset; (b) has $\delta t=1,M=2,m_\ins=6$ and varying $m_\out$; (c) has $m_\ins=1,m_\out=2,M=2$ and varying $\delta t$; and (d) was made with $m_\ins=1,m_\out=2,\delta t=1$ while $M$ was changed.}
	\label{fig:cpxt}
\end{figure}

The goal is to understand how the dynamics of $\Delta C_{F_2}(t)$ is affected by the quench parameters $\delta t,m_\ins,m_\out$ (or $\delta m\equiv m_\out-m_\ins$), which characterize the target state $\ket{\psi(t)}$, as well as by the scale $M$ that characterizes the reference state $\ket\Omega$. Notice that, with the mass taken to depend on time as defined in \eq{massquenchtanh}, the quench effectively starts at $t \approx -\delta t$ and ends at $t\approx \delta t$, so the plots below will always focus on this interval $(-\delta t, \delta t$) where a nontrivial time evolution is expected. The relevant results are shown in Figure \ref{fig:cpxt}. For numerical convenience we have fixed $\Lambda=100$  and all the plots were made using specific numerical values for the parameters chosen for presentation purposes only, although the qualitative behavior illustrated in each one has been checked to hold generically.

Regardless of the choice of parameters one can see two distinct regimes for the time dependence of the complexity that are illustrated in Figures (a) through (d): an approximate linear evolution for $-\delta t\lesssim t \lesssim \delta t$, followed by a saturation phase at $t \gtrsim \delta t$, where the complexity approaches a constant with eventual fluctuations around this value. As illustrated in (a) and (b), the slope of the linear phase can be positive or negative (i.e., the complexity can grow or shrink) depending on the sign of $\delta m$. Namely, for quenches that increase the mass ($\delta m >0$) the slope is always positive, while for mass-decreasing quenches it is negative. The amplitude of the final state oscillations is influenced not only by the mass difference $\delta m$, but also by the initial value $m_\ins$ as one can see by comparing (a) and (b): the former has $m_\ins=1$ and the oscillations become more pronounced as $\delta m$ is increased, while the latter with $m_\ins=6$ and decreasing $\delta m$ shows almost imperceptible oscillations. 

The quench rate $\delta t$ also has significant influence in the time evolution as indicated in (c). First, the slope of the linear phase increases the faster the quench is done. Second, the late time oscillations of the complexity are also sensitive to $\delta t$ and become more pronounced as $\delta t$ decreases. This is compatible with the infinitely fast quench limit $\delta t\to0$ discussed analytically in Section \ref{sec:fast}, which is also shown in the plot, where the complexity jumps (hence having an infinite slope) at $t=0$ from a constant value to an oscillatory behavior for $t>0$.

In (d) we analyze also the dependence of the complexity of the energy scale $M$ that defines the reference state. To do this, we keep the quench data $m_\ins,m_\out,\delta t$ (i.e., the target state) fixed. Interestingly, we see that the renormalized complexity growth rate decreases and becomes even negative as the scale $M$ is increased. In particular, there is a particular scale ($M\approx3$ for the set of parameters used in the plot) for which there is a transition from growth to decrease of the renormalized complexity. At this scale, $\Delta C_{F_2}(t)$ remains approximately zero during the whole time evolution, i.e., the quench barely increases the complexity of the initial state. It is important to keep in mind that varying $M$ affects not only the instantaneous complexity $C_{F_2}(t)$, but also the vacuum contribution $C_{F_2}^\text{vacuum}$ that is being subtracted to yield $\Delta C_{F_2}(t)$.

	\subsection{Higher dimensions}
    
The analysis of the complexity above is easily extended to an arbitrary number of spacetime dimensions. Namely, the gate operators now carry a vector index $\kk$, i.e. $\Op_{I,\kk}$, but still close the same SU($1,1$) algebra. The reference and target states remain unchanged, since the out modes $\chi^\out_k(t)$ are the same in higher dimensions (see e.g. \cite{Das:2014hqa}) and therefore the defining constraints \eq{reftargetdef} do not care about the number of dimensions. The circuit is thus essentially the same and the calculation of geodesics proceeds in a similar way by focusing on a single $\kk$ to later join them all together. As a result, the complexity in $d=D+1$ dimensions takes the same form as in \eq{cpxt_finalanswer} but with $$\int_{-\Lambda}^{\Lambda} dk \rightarrow \int d^D\kk = S_{D-1}\int_0^\Lambda dk\,k^{D-1}\,,$$
where $S_{D-1}$ is the area of the ($D-1$)-sphere.\footnote{Of course there is also the spatial volume factor $V$ in front of the momentum integrals that we are here ignoring and setting to $1$.}

\section{Discussion and Conclusion}\label{conclusion}

In this work, we have studied the time evolution of the computational complexity as the result of a smooth mass quench in a free bosonic field theory. This model at late times is known to thermalize in the sense of generalized Gibbs ensemble \cite{Sotiriadis:2014uza,Alves:2017fjk}, so this can be seen as a toy model for studying the evolution of complexity during a thermalization process. The reference state was chosen to be one with no spatial entanglement, while the target state was a time-dependent one obtained from the pre-quench vacuum through time evolution. As elementary gates, we chose the ones appearing in the momentum-space Hamiltonian \eq{Hamiltonian} of the model. Since they close a SU($1,1$) algebra, one can follow the procedure of Jefferson and Myers in \cite{Jefferson:2017sdb} to geometrize the problem of finding the complexity as the one of finding a geodesic in the manifold of SU($1,1$) transformations. 

The main result of the paper is formula \eq{cpxt_finalanswer} for the time evolution of the complexity $C_{F_2}(t)$ in terms of the data ($A_k,B_k,\alpha_k(t),\beta_k(t)$) characterizing the reference and target states. In the limit where there is no quench and we have simply a massive scalar field in the vacuum state, the time dependence cancels out and we recover the same expression \eq{cpxtvacuum} for the complexity obtained in \cite{Jefferson:2017sdb}. This is interesting since the set of elementary gates chosen here is quite different from the ones used in that case. 

The complexity itself is well-known to be UV divergent, since there are infinitely many momentum modes (gates) $k$ contributing to the circuit. However, we have shown that the UV divergent piece happens to be time-independent and exactly the same as the one from the initial state complexity $C_{F_2}^\text{vacuum}$, such that the difference $\Delta C_{F_2}(t)\equiv C_{F_2}(t)-C_{F_2}^\text{vacuum}$ (the \lq\lq renormalized\rq\rq~complexity) is by construction UV finite. We then studied the time dependence of this object as a function of the quench data $m_\ins,m_\out,\delta t$ as well as the free scale $M$ characterizing the reference state.

We found that the time evolution of the renormalized complexity is marked by two distinct phases, separated by a time scale set by the quench rate $\delta t$ (or, equivalently, the thermalization scale for the model). For $-\delta t\lesssim t\lesssim\delta t$, the complexity grows (or decays, depending on the sign of $\delta m\equiv m_\out-m_\ins$) approximately in a linear way, with a slope $\frac{dC}{dt}$ inversely proportional to $\delta t$ and directly proportional to $\delta m$. Then, at a time of order $\delta t$ the complexity saturates and starts oscillating around a mean value. We checked that the amplitude of the late time oscillations is sensitive not only to $\delta t$ and the mass difference $\delta m$, but also to the starting point of the quench ($m_\ins$) and the reference scale $M$. Also, for fixed quench data the renormalized complexity shifts from growing to decreasing in time as the scale $M$ is increased, and in particular there is an intermediate scale for which the change in complexity due to the quench is nearly vanishing.

As mentioned before, it has been conjectured that the complexity growth might be dual to the growth of Einstein-Rosen bridges in eternal AdS black hole geometries. Classical gravity computations show that observables characterizing this growth \cite{Carmi:2017jqz} evolve linearly at late times. The holographic complexity was observed to grow linearly as well in AdS-Vaidya model of a quantum quench \cite{Moosa:2017yvt}. Also, a linear growth followed by saturation with oscillations around a mean value had been previously conjectured to be the universal behavior of the complexity for chaotic systems \cite{Brown:2017jil}, \cite{Brown:2016wib}. The time scale for saturation in these systems is believed to be parametrically larger than the thermalization scale. Our results show some similarities with these conjectures and calculations even though we dealt with a free (hence non-chaotic) field theory, with the important difference that the time scale for saturation here is of the order of the thermalization scale $\delta t$. 
For this reason, it seems to be an interesting future prospect to extend the calculations presented here to interacting theories, which are the ones most likely to have a simple classical gravitational dual, and check if the early linear behavior shown here can be pushed to last up to larger time scales. Another interesting prospect in this direction would be to use the $\kappa=1$ measure discussed in \cite{Jefferson:2017sdb} instead of the $F_2$-distance used here, which has been argued to be more appropriate to get closer to holographic results.


\subsection*{Acknowledgements}

We would like to thank Rob Myers and Veronika Hubeny for valuable comments on the draft, and also Ying Zhao, and Mukund Rangamani for discussions. 
D.W.F.A. would like to acknowledge hospitality at the QMAP center of UC Davis, where part of this work was developed. 
D.W.F.A is supported  by the PDSE Program scholarship of CAPES – Brazilian Federal Agency for Support and Evaluation of Graduate Education within the Ministry of Education of Brazil, and by the CNPq grant 146086/2015-5.
G.C. acknowledges financial support from the Brazilian ministries MCTI and MEC.

\appendix
\section{Relation between Schr\"odinger and Heisenberg picture circuits}
\label{app:SH}

\setcounter{equation}{0}
\renewcommand{\theequation}{\thesection.\arabic{equation}}

As introduced in Section \ref{introducing_the_circuit}, in the Schr\"odinger picture the target state is time-dependent while the reference state is not. Namely, the target state is $\ket{\psi\left(t\right)}$ obtained by evolving on time the initial state $\ket{0_{\ins}}$ while the reference state $\ket{\Omega}$ is constrained to satisfy 
\begin{equation}\label{eq:refstateSch}
\left(\sqrt{M}\,\phi\left(x,0\right)+\frac{\ii}{\sqrt{M}}\,\pi\left(x,0\right)\right)\ket{\Omega}=0\,.
\end{equation}
where we have chosen, with no loss of generality, the time slice $t=0$ for the Schr\"odinger picture operators. The optimal circuit is 
\begin{equation}
\ket{\psi\left(t\right)} = U_\text{optimal}^\text{Sch}\left(s=1,t\right)\ket{\Omega}\equiv \mathcal{P}\,e^{\ii\int_{0}^{1}ds'\,Y^{I}\left(t,s'\right)\Op_{I}}\,\ket{\Omega}\,,
\end{equation}
which can also be written as 
\begin{equation}
\ket{\psi\left(0\right)} = E\left(t,0\right)^{-1}\,\mathcal{P}\,e^{\ii\int_{0}^{1}ds'\,Y^{I}\left(t,s'\right)\Op_{I}}\ket{\Omega}\,,
\end{equation}
where $E\left(t,0\right)$ is the time evolution operator. In the Schr\"odinger picture, the task is therefore to find a circuit as above satisfying the boundary condition
\begin{equation}
U_\text{optimal}^\text{Sch}\left(s=0,t\right)=1\,,
\end{equation}
which amounts to finding the optimal choice of $Y^{I}\left(t,s\right)$ that minimizes $\int ds\,\sqrt{\delta_{IJ}Y^{I}Y^{J*}}$.

In the Heisenberg picture, on the other hand, the target state is time-independent while the reference
state is not. The target reads
\begin{equation}
\ket{\psi\left(0\right)}\equiv E\left(t,0\right)^{-1}\ket{\psi\left(t\right)}
\end{equation}
while the reference state is given by
\begin{equation}
\ket{\Omega\left(t\right)} \equiv E\left(t,0\right)^{-1}\ket\Omega\,.
\end{equation}
By acting with $E\left(t,0\right)^{-1}$ on \eq{refstateSch} it is straightforward to see that such a reference state satisfies a similar relation for the Heisenberg picture quantities, namely 
\begin{equation}
\left(\sqrt{M}\,\phi\left(x,t\right)+\frac{\ii}{\sqrt{M}}\,\pi\left(x,t\right)\right)\ket{\Omega\left(t\right)}=0\,.
\end{equation}
One can write the optimal circuit as 
\begin{eqnarray}
\ket{\psi\left(0\right)} &=& U_\text{optimal}^\text{Hei}\left(s=1,t\right)\ket{\Omega\left(t\right)}\notag\\
&=& \mathcal{P}\,e^{\ii\int_{0}^{1}ds'\,X^{I}\left(t,s'\right)\Op_{I}}\ket{\Omega\left(t\right)}\notag\\
&=&\mathcal{P}\,e^{\ii\int_{0}^{1}ds\,X^{I}\left(t,s\right)\Op_{I}}E\left(t,0\right)^{-1}\ket\Omega\,.
\end{eqnarray}
and the task is to find the set of functions $X^{I}$ that satisfy the boundary condition
\begin{equation}
U_\text{optimal}^\text{Hei}\left(s=0,t\right)=1
\end{equation}
and minimize $\int ds\,\sqrt{\delta_{IJ}X^{I}X^{J*}}$. 

The question then is how to relate the two apparently distinct ways to calculate the complexity. To answer that, let us multiply the last condition by $E^{-1}E=1$ on the left to obtain 
\begin{equation}
\ket{\psi\left(0\right)} = E^{-1}\left(t,0\right)\left(E\left(t,0\right)\mathcal{P}\,e^{\ii\int_{0}^{1}ds\,X^{I}\left(t,s\right)\Op_{I}}E^{-1}\left(t,0\right)\right)\ket{\Omega}\,.
\end{equation}
Since $E$ does not depend on $s$ (the circuit \lq\lq level\rq\rq), it can go inside the path-ordering such that
\begin{equation}\label{eq:heisenberg_condition}
\ket{\psi\left(0\right)} = E\left(t,0\right)^{-1}\left(\mathcal{P}\,e^{\ii\int_{0}^{1}ds\,X^{I}\left(t,s\right)\left[E\left(t,0\right)\Op_{I}E\left(t,0\right)^{-1}\right]}\right)\ket{\Omega}\,.
\end{equation}


Now, consider the Heisenberg picture version of the operator $\Op_I$, namely,
\begin{equation}\label{eq:HeisSchOp}
O_{I}\left(t\right)\equiv E\left(t,0\right)^{-1}O_{I}\,E\left(t,0\right)\,,
\end{equation}
which satisfies
\begin{equation}
\frac{\partial O_{I}\left(t\right)}{\partial t} = \frac{\partial}{\partial t}\left(\overline{\mathcal{T}}\,e^{\ii\int_{0}^{t}dt'\,H\left(t'\right)}\Op_{I}\,\mathcal{T}\,e^{-\ii\int_{0}^{t}dt'\,H\left(t'\right)}\right) 
= \ii E\left(t,0\right)^{-1}\left[H\left(t\right),O_{I}\right] E\left(t,0\right)\,.
\end{equation}
Recall that the Hamiltonian for our problem is a combination of creation and annihilation operators that close a SU($1,1$) algebra, and we have chosen the $\{O_{I}\}$ to be exactly that set of operators, that is,
\begin{equation}
H\left(t\right)=f^{J}\left(t\right)O_{J}\,
\end{equation}
for some functions $f^J(t)$ that can be seen from \eq{Hamiltonian}.
This means that 
\begin{equation}
\frac{\partial O_{I}\left(t\right)}{\partial t}
=\ii f^{J}\left(t\right)c_{JIK}O_{K} (t)\,,
\end{equation}
where $c_{JIK}$ are the structure constants of SU($1,1$). 
The solution is 
\begin{equation}\label{eq:aux}
O_{I}\left(t\right)=M_{I}^{\,\,J}\left(t\right)O_{J}
\end{equation}
where $M_{I}$ is an element of SU($1,1$) in the adjoint representation.

We can now return to \eq{heisenberg_condition} and write
\begin{equation}
\ket{\psi\left(0\right)}=E\left(t,0\right)^{-1}\left(\mathcal{P}\,e^{\ii\int_{0}^{1}ds\,X^{I}\left(t,s\right)\left(M^{-1}\right)_{IK}M^{KL}\left[E\left(t,0\right)\Op_{L}E\left(t,0\right)^{-1}\right]}\right)\ket{\Omega}\,.
\end{equation}
Defining $\tilde{X}_{K}\equiv X^{I}\left(M^{-1}\right)_{IK}$ and using \eq{aux} together with \eq{HeisSchOp} to identify the Schrödinger picture operator $\Op_I$ in the exponent, one can now rephrase the Heisenberg problem as that of finding the circuit 
\begin{equation}
\ket{\psi\left(0\right)} = E\left(t,0\right)^{-1}\left(\mathcal{P}\,e^{\ii\int_{0}^{1}ds\,\tilde{X}^{k}\left(t,s\right)O_{K}}\right)\ket{\Omega}
\end{equation}
that minimizes 
\begin{equation}
\int ds\,\sqrt{\delta_{IJ}X^{I}X^{J*}} = \int ds\sqrt{(M)^I_K\tilde{X}^{K}\delta_{IJ}(M)^{J*}_L\tilde{X}^{L*}} = \int ds\sqrt{(M)^I_K\tilde{X}^{K}(M)^{I*}_L\tilde{X}^{L*}}
\end{equation}
Finally, by noticing that $(M)^I_K\,(M)^{I*}_L = (MM^{\dagger})_{KL}=\delta_{KL}$ since $M\in$ SU($1,1$) we see that this becomes exactly the Schr\"odinger picture problem, i.e., that of finding the circuit
\begin{equation}
\ket{\psi\left(0\right)} = E\left(t,0\right)^{-1}\left(\mathcal{P}\,e^{\ii\int_{0}^{1}ds\,\tilde{X}^{k}\left(t,s\right)\Op_{k}}\right)\ket{\Omega}
\end{equation}
that minimizes 
\begin{equation}
\int ds\sqrt{\delta^{KL}\tilde{X}_{K}\tilde{X}_{L}^{*}}\,.
\end{equation}

\bibliographystyle{JHEP}
\bibliography{refs_cpxt}

\end{document}